\documentclass[preprint,aps,prc,superscriptaddress,showpacs,floatfix]{revtex4}
\usepackage{amssymb}
\usepackage{amsmath}
\usepackage{graphicx}
\usepackage{longtable}
\usepackage[normalem]{ulem}
\usepackage[dvips]{color}

\renewcommand\sout{\bgroup \color{red} \ULdepth=-.5ex \ULset}

\begin{document}

\title{Higher-order effects on the incompressibility of isospin asymmetric nuclear matter}
\author{Lie-Wen Chen}
\affiliation{Department of Physics, Shanghai Jiao Tong University, Shanghai 200240, China}
\affiliation{Center of Theoretical Nuclear Physics, National Laboratory of Heavy Ion
Accelerator, Lanzhou 730000, China}
\author{Bao-Jun Cai}
\affiliation{Department of Physics, Shanghai Jiao Tong University, Shanghai 200240, China}
\author{Che Ming Ko}
\affiliation{Cyclotron Institute and Physics Department, Texas A\&M University, College
Station, Texas 77843-3366, USA}
\author{Bao-An Li}
\affiliation{Department of Physics, Texas A\&M University-Commerce, Commerce, Texas
75429-3011, USA}
\author{Chun Shen}
\affiliation{Department of Physics, Shanghai Jiao Tong University, Shanghai 200240, China}
\author{Jun Xu}
\affiliation{Cyclotron Institute and Physics Department, Texas A\&M University, College
Station, Texas 77843-3366, USA}
\date{\today }

\begin{abstract}
Analytical expressions for the saturation density of asymmetric
nuclear matter as well as its binding energy and incompressibility
at saturation density are given up to $4$th-order in the isospin
asymmetry $\delta =(\rho _{n}-\rho _{p})/\rho $ using $11$
characteristic parameters defined by the density derivatives of the
binding energy per nucleon of symmetric nuclear matter, the symmetry
energy $E_{\text{\textrm{sym}}}(\rho )$ and the $4$th-order symmetry
energy $E_{\text{\textrm{sym,4}}}(\rho )$ at the normal nuclear
density $\rho _{0}$. Using an isospin- and momentum-dependent
modified Gogny (MDI) interaction and the Skyrme-Hartree-Fock (SHF)
approach with $63$ popular Skyrme interactions, we have
systematically studied the isospin dependence of the saturation
properties of asymmetric nuclear matter, particularly the
incompressibility $K_{\mathrm{sat}}(\delta
)=K_{0}+K_{\mathrm{sat,2}}\delta ^{2}+K_{\mathrm{sat,4}}\delta
^{4}+O(\delta ^{6})$ at saturation density. Our results show that
the magnitude of the higher-order $K_{\mathrm{sat,4}}$ parameter is
generally small compared to that of the $K_{\mathrm{sat,2}}$
parameter. The latter essentially characterizes the isospin
dependence of the incompressibility at saturation density and can be
expressed as
$K_{\mathrm{sat,2}}=K_{\mathrm{sym}}-6L-\frac{J_{0}}{K_{0}}L$, where
$L$ and $K_{\mathrm{sym}}$ represent, respectively, the slope and
curvature parameters of the symmetry energy at $\rho _{0}$ while
$J_{0}$ is the third-order derivative parameter of symmetric nuclear
matter at $\rho _{0}$. Furthermore, we have constructed a
phenomenological modified Skyrme-like (MSL) model which can
reasonably describe the general properties of symmetric nuclear
matter and the symmetry energy predicted by both the MDI model and
the SHF approach. The results indicate that the higher-order $J_{0}$
contribution to $K_{\mathrm{sat,2}}$ generally cannot be neglected.
In addition, it is found that there exists a nicely linear
correlation between $K_{\mathrm{sym}}$ and $L$ as well as between
$J_{0}/K_{0}$ and $K_{0}$. These correlations together with the
empirical constraints on $K_{0}$, $L$, $E_{\text{\textrm{sym}}}(\rho
_{0})$ and the nucleon effective mass lead to an estimate of
$K_{\mathrm{sat,2}}=-370\pm 120$ MeV.
\end{abstract}

\maketitle

\section{Introduction}

The equation of state (EOS) of nuclear matter is one of central
issues in nuclear physics. For a cold nuclear matter, the EOS is
usually defined as the binding energy per nucleon as a function of
the density from which information on other thermodynamic properties
of nuclear matter, such as its pressure and incompressibility can be
obtained. With the establishment or construction of many radioactive
beam facilities around the world, such as the Cooling Storage Ring
(CSR) facility at HIRFL in China \cite{CSR}, the Radioactive Ion
Beam (RIB) Factory at RIKEN in Japan \cite{Yan07}, the FAIR/GSI in
Germany \cite{FAIR}, SPIRAL2/GANIL in France \cite{SPIRAL2}, and the
Facility for Rare Isotope Beams (FRIB) in USA \cite{RIA}, it is
possible in terrestrial laboratories to explore the EOS of an
isospin asymmetric nuclear matter under the extreme condition of
large isospin asymmetry, especially the density dependence of the
nuclear symmetry energy. This knowledge, especially the latter, is
important for understanding not only the structure of radioactive
nuclei, the reaction dynamics induced by rare isotopes, and the
liquid-gas phase transition in asymmetric nuclear
matter, but also many critical issues in astrophysics \cite%
{LiBA98,LiBA01b,Dan02a,Lat00,Lat01,Lat04,Bar05,Ste05a,CKLY07,LCK08}.

For symmetric nuclear matter with equal fractions of neutrons and
protons, its EOS is relatively well-determined after about more than
$30$ years of studies by the nuclear physics community. In
particular, the incompressibility of symmetric nuclear matter at its
saturation density $\rho _{0}$ has been determined to be $240\pm 20$
MeV from analyses of the nuclear giant monopole resonances (GMR)
\cite{You99,Lui04,Ma02,Vre03,Col04,Shl06,LiT07,Gar07,Col09}, and its
EOS at densities of $2\rho _{0}<\rho <5\rho _{0}$ has also been
constrained by measurements of collective flows in nucleus-nucleus
collisions \cite{Dan02a} and of subthreshold kaon production
\cite{Aic85,Fuc06a} in relativistic nucleus--nucleus collisions. On
the other hand, the EOS of asymmetric nuclear matter, especially the
density dependence of the nuclear symmetry energy, is largely
unknown. Although the nuclear symmetry energy at $\rho _{0}$ is
known to be around $30$ MeV from the empirical liquid-drop mass
formula \cite{Mey66,Pom03}, its values at other densities,
especially at supra-saturation densities, are poorly known
\cite{LiBA98,LiBA01b}. Various microscopic and phenomenological
models, such as the relativistic
Dirac-Brueckner-Hartree-Fock (DBHF) \cite%
{Ulr97,Fuc04,Ma04,Sam05a,Fuc05,Fuc05b,Ron06} and the non-relativistic
Brueckner-Hartree-Fock (BHF) \cite{Bom91,Zuo05,LiZH06,Bal07} approach, the
relativistic mean-field (RMF) model based on nucleon-meson interactions \cite%
{Ren02,Bar05,Men06,Che07}, and the non-relativistic mean-field model based
on Skyrme-like interactions \cite%
{Das03,LiBA04a,LiBA04c,Che04,Riz04,Beh05,Riz05,Che05b,Che09}, have
been used to study the isospin-dependent properties of asymmetric
nuclear matter, such as the nuclear symmetry energy, the nuclear
symmetry potential, the isospin-splitting of the nucleon effective
masses, etc., but the predicted results vary widely. In fact, even
the sign of the symmetry energy above $3\rho _{0}$ is still
uncertain \cite{Bom01,Kub03}. The theoretical uncertainties are
mainly due to the lack of knowledge about the isospin dependence of
in-medium nuclear effective interactions and the limitations in the
techniques for solving the nuclear many-body problem.

Heavy-ion collisions, especially those induced by neutron-rich
nuclei, provide a unique tool to investigate the EOS of asymmetric
nuclear matter, especially the density dependence of the nuclear
symmetry energy. During the last decade, significant progress has
indeed been made both experimentally and theoretically on
constraining the behavior of the symmetry energy at subsaturation
density using heavy-ion reactions \cite%
{Tsa04,Che05a,Che05b,LiBA05c,She07,Zha08,Tsa09,Tra09} (See Ref.
\cite{LCK08} for the most recent review). More recently, the IBUU04
transport model analysis of the FOPI data on the $\pi ^{-}/\pi ^{+}$
ratio in central heavy-ion collisions at SIS/GSI \cite{Rei07}
energies suggests a very soft symmetry energy at the suprasaturation
densities \cite{Xiao09}. Information on the density dependence of
the nuclear symmetry energy has also been obtained from the
structure of finite nuclei and their excitations, such as the mass
data \cite{Mye96}, neutron skin in heavy nuclei \cite{Bro00}, giant
dipole resonances \cite{Tri08}, pygmy dipole resonance \cite{Kli07},
and so on. These studies have significantly improved our
understanding of the EOS of asymmetric nuclear matter.

The incompressibility of asymmetric nuclear matter at its saturation
density is a basic quantity to characterize its EOS. Since this
quantity is largely undetermined, any constraint imposed on it would
be extremely important. In the present work, we study the isospin
dependence of the properties of asymmetric nuclear matter, including
the saturation density as well as the binding energy and
incompressibility at saturation density. In particular, we derive
analytical expressions for these quantities up to the $4$th-order in
the isospin asymmetry $\delta =(\rho _{n}-\rho _{p})/\rho $ and
investigate the higher-order isospin asymmetry effects on the
properties of asymmetric nuclear matter. For the incompressibility
of an asymmetric nuclear matter at its saturation density, it can be
written as $K_{\mathrm{sat}}(\delta )=K_{0}+K_{\mathrm{sat,2}}\delta
^{2}+K_{\mathrm{sat,4}}\delta ^{4}+O(\delta ^{6})$ with
$K_{\mathrm{sat,2}}=K_{\mathrm{sym}}-6L-\frac{J_{0}}{K_{0}}L$, where
$L$ and $K_{\mathrm{sym}}$ represent, respectively, the slope and
curvature parameters of the symmetry energy at $\rho _{0}$ while
$J_{0}$ is the $3$rd-order derivative parameter of symmetric nuclear
matter at $\rho _{0}$. Therefore, the higher-order effects on
$K_{\mathrm{sat}}(\delta )$ also include the $3$rd-order density
derivative parameter $J_{0}$. Our results indicate that higher-order
isospin asymmetry effects on the incompressibility are usually not
important but the higher-order $J_{0}$ contribution generally cannot
be neglected. Furthermore, we construct a phenomenological modified
Skyrme-like (MSL) model which can reasonably describe the general
properties of symmetric nuclear matter and the symmetry energy
predicted by both the MDI model and the Skyrme-Hartree-Fock (SHF)
approach. We find that there exists a nicely linear correlation
between $K_{\mathrm{sym}}$ and $L$ as well as between $J_{0}/K_{0}$
and $K_{0}$. These correlations together with the empirical
constraints on $K_{0}$, $L$ and $E_{ \text{\textrm{sym}}}(\rho
_{0})$ lead to an estimate of $K_{\mathrm{sat,2} }=-370\pm 120$ MeV.

The paper is organized as follows. In Section \ref{saturation}, we
discuss the general properties of asymmetric nuclear matter, and
then give analytical expressions up to $4$th-order terms in isospin
asymmetry $\delta $ for the saturation density of asymmetric nuclear
matter as well as its binding energy and incompressibility at
saturation density.We then briefly introduce in Section \ref{models}
the three models used in the present paper, i.e., the isospin- and
momentum-dependent MDI model, the SHF approach, and the
phenomenological MSL model. The results and discussions are
presented in Section \ref{results}. A summary is then given in
Section \ref{summary}. For completeness, derivations of some
important formula shown in Section \ref{saturation} are briefly
described in the Appendix.

\section{Saturation properties of asymmetric nuclear matter}

\label{saturation}

\subsection{Equation of state of asymmetric nuclear matter}

The EOS of isospin asymmetric nuclear matter, given by its binding
energy per nucleon, can be generally expressed as a power series in
the isospin asymmetry $\delta =(\rho _{n}-\rho _{p})/\rho$, where
$\rho =\rho _{n}+\rho _{p}$ is the baryon density with $\rho _{n}$
and $\rho _{p}$ denoting the neutron and proton densities,
respectively. To the $4$th-order in isospin asymmetry, it is written
as
\begin{equation}
E(\rho ,\delta )=E_{0}(\rho )+E_{\mathrm{sym}}(\rho )\delta
^{2}+E_{\mathrm{sym,4}}(\rho )\delta ^{4}+O(\delta ^{6}),
\label{EOSANM}
\end{equation}%
where $E_{0}(\rho )=E(\rho ,\delta =0)$ is the binding energy per
nucleon of symmetric nuclear matter, and
\begin{eqnarray}
E_{\mathrm{sym}}(\rho ) &=&\frac{1}{2!}\frac{\partial ^{2}E(\rho ,\delta )}
{\partial \delta ^{2}}|_{\delta =0}\label{Esym} \\
E_{\mathrm{sym,4}}(\rho ) &=&\frac{1}{4!}\frac{\partial ^{4}E(\rho
,\delta ) }{\partial \delta ^{4}}|_{\delta =0}.  \label{Esym4}
\end{eqnarray}%
In the above, $E_{\mathrm{sym}}(\rho )$ is the\ so-called nuclear
symmetry energy and $E_{\mathrm{sym,4}}(\rho )$ is the $4$th-order
coefficient, which is called here the $4$th-order nuclear symmetry
energy. The absence of odd-order terms in $\delta $ in Eq.
(\ref{EOSANM}) is due to the exchange symmetry between protons and
neutrons in nuclear matter when one neglects the Coulomb interaction
and assumes the charge symmetry of nuclear forces. The nuclear
symmetry energy $E_{\mathrm{sym}}(\rho )$ thus corresponds to the
lowest-order coefficient. The higher-order (including $4$th-order)
coefficients in $\delta $ are usually very small and neglected,
e.g., the magnitude of the $\delta ^{4}$ term at the normal nuclear
density $\rho _{0}$ (the saturation density of symmetric nuclear
matter) is estimated to be less than $1$ MeV in microscopic
many-body approaches \cite{Sie70,Sjo74,Lag81} and also in
phenomenological models as shown later. Neglecting the contribution
from higher-order terms in Eq. (\ref{EOSANM}) leads to the
well-known empirical parabolic law for the EOS of asymmetric nuclear
matter, which has been verified by all many-body theories to date,
at least for densities up to moderate values \cite{LCK08}. As a good
approximation, the density-dependent symmetry energy
$E_{\mathrm{sym}}(\rho ) $ can thus be extracted from the parabolic
approximation of $E_{\mathrm{sym}}(\rho )\approx E(\rho ,\delta
=1)-E(\rho ,\delta =0)$. It should be mentioned that the possible
presence of the higher-order terms in $\delta $ at supra-normal
densities can significantly modify the proton fraction in $\beta
$-equilibrium neutron-star matter and the critical density for the
direct Urca process which can lead to faster cooling of neutron
stars \cite{Zha01,Ste06}. In addition, a recent study
\cite{Xu09a,Xu09b} indicates that the higher-order terms in $\delta
$ are very important for determining the transition density and
pressure at the inner edge separating the liquid core from the solid
crust of neutron stars where the matter is extremely neutron-rich.

Around the nuclear matter saturation density $\rho _{0}$, the
binding energy per nucleon in symmetric nuclear matter $E_{0}(\rho
)$\ can be expanded, e.g., up to $4$th-order in density, as
\begin{eqnarray}
E_{0}(\rho ) &=&E_{0}(\rho _{0})+L_{0}\chi +\frac{K_{0}}{2!}\chi ^{2}  \notag
\\
&&+\frac{J_{0}}{3!}\chi ^{3}+\frac{I_{0}}{4!}\chi ^{4}+O(\chi ^{5}),
\label{E0}
\end{eqnarray}%
where $\chi $ is a dimensionless variable characterizing the
deviations of the density from the saturation density $\rho _{0}$ of
symmetric nuclear matter and it is conventionally defined as

\begin{equation}
\chi =\frac{\rho -\rho _{0}}{3\rho _{0}}.  \label{chi}
\end{equation}%
The first term $E_{0}(\rho _{0})$ on the right-hand-side (r.h.s) of
Eq. (\ref{E0}) is the binding energy per nucleon in symmetric
nuclear matter at the saturation density $\rho _{0}$ and the
coefficients of other terms are
\begin{eqnarray}
L_{0} &=&3\rho _{0}\frac{dE_{0}(\rho )}{d\rho }|_{\rho =\rho _{0}}, \\
K_{0} &=&9\rho _{0}^{2}\frac{d^{2}E_{0}(\rho )}{d\rho ^{2}}|_{\rho =\rho
_{0}}, \\
J_{0} &=&27\rho _{0}^{3}\frac{d^{3}E_{0}(\rho )}{d\rho ^{3}}|_{\rho =\rho
_{0}}, \\
I_{0} &=&81\rho _{0}^{4}\frac{d^{4}E_{0}(\rho )}{d\rho ^{4}}|_{\rho =\rho
_{0}}.
\end{eqnarray}%
Obviously, we have $L_{0}=0$ according to the definition of the
saturation density $\rho _{0}$ of symmetric nuclear matter and thus
the second term on the r.h.s of Eq. (\ref{E0}) should vanish. The
coefficient $K_{0}$ is the so-called incompressibility coefficient
of symmetric nuclear matter and it characterizes the curvature of
$E_{0}(\rho )$ at $\rho _{0}$. The coefficients $J_{0}$ and $I_{0}$
correspond to higher-order contributions and here we call them as
$3$rd-order and $4$th-order incompressibility coefficients of
symmetric nuclear matter, respectively. In the literature, one
usually neglects the higher-order terms in Eq. (\ref{E0}) around the
saturation density $\rho _{0}$ and obtain the following parabolic
approximation to the EOS of symmetric nuclear matter:
\begin{equation}
E_{0}(\rho )=E_{0}(\rho _{0})+\frac{K_{0}}{2}\chi ^{2}+O(\chi ^{3}).
\label{E0para}
\end{equation}

Around the normal nuclear density $\rho _{0}$, the nuclear symmetry
energy $E_{\mathrm{sym}}(\rho )$\ can be similarly expanded, e.g.,
up to $4$th-order in $\chi $, as
\begin{eqnarray}
E_{\mathrm{sym}}(\rho ) &=&E_{\mathrm{sym}}(\rho _{0})+L\chi
+\frac{K_{\mathrm{sym}}}{2!}\chi ^{2}  \notag \\
&&+\frac{J_{\mathrm{sym}}}{3!}\chi
^{3}+\frac{I_{\mathrm{sym}}}{4!}\chi^{4}+O(\chi ^{5}),
\label{EsymLKJI}
\end{eqnarray}%
where $L$, $K_{\mathrm{sym}}$, $J_{\mathrm{sym}}$ and
$I_{\mathrm{sym}}$ are the slope parameter, curvature parameter,
$3$rd-order coefficient, and $4$th-order coefficient of the nuclear
symmetry energy at $\rho _{0}$, i.e.,
\begin{eqnarray}
L &=&3\rho _{0}\frac{dE_{\mathrm{sym}}(\rho )}{\partial \rho }|_{\rho =\rho
_{0}},  \label{L} \\
K_{\mathrm{sym}} &=&9\rho _{0}^{2}\frac{d^{2}E_{\mathrm{sym}}(\rho )}
{\partial \rho ^{2}}|_{\rho =\rho _{0}},  \label{Ksym} \\
J_{\mathrm{sym}} &=&27\rho _{0}^{3}\frac{d^{3}E_{\mathrm{sym}}(\rho )}
{\partial \rho ^{3}}|_{\rho =\rho _{0}}, \\
I_{\mathrm{sym}} &=&81\rho _{0}^{4}\frac{d^{4}E_{\mathrm{sym}}(\rho
)}{\partial \rho ^{4}}|_{\rho =\rho _{0}}.
\end{eqnarray}%
The coefficients $L$, $K_{\mathrm{sym}}$, $J_{\mathrm{sym}}$ and
$I_{\mathrm{sym}}$ characterize the density dependence of the
nuclear symmetry energy around the normal nuclear density $\rho
_{0}$, and thus carry important information on the properties of
nuclear symmetry energy at both high and low densities.

We can also expand the $4$th-order nuclear symmetry energy
$E_{\mathrm{sym,4}}(\rho )$ around the normal nuclear density $\rho
_{0}$ up to $4$th-order in $\chi $ as
\begin{eqnarray}
E_{\mathrm{sym,4}}(\rho ) &=&E_{\mathrm{sym,4}}(\rho _{0})+L_{\mathrm{sym,4}}
\chi +\frac{K_{\mathrm{sym,4}}}{2}\chi ^{2}  \notag \\
&&+\frac{J_{\mathrm{sym,4}}}{3!}\chi
^{3}+\frac{I_{\mathrm{sym,4}}}{4!}\chi^{4}+O(\chi ^{5}),
\label{Esym4LKJI}
\end{eqnarray}%
where $L_{\mathrm{sym,4}}$, $K_{\mathrm{sym,4}}$,
$J_{\mathrm{sym,4}}$ and $I_{\mathrm{sym,4}}$ are the slope
parameter, curvature parameter, $3$rd-order coefficient, and
$4$th-order coefficient of the $4$th-order nuclear symmetry energy
$E_{\mathrm{sym,4}}(\rho )$ at $\rho _{0}$, i.e.,
\begin{eqnarray}
L_{\mathrm{sym,4}} &=&3\rho _{0}\frac{dE_{\mathrm{sym,4}}(\rho )}{d\rho }
|_{\rho =\rho _{0}}, \\
K_{\mathrm{sym,4}} &=&9\rho _{0}^{2}\frac{d^{2}E_{\mathrm{sym,4}}(\rho )}
{d\rho ^{2}}|_{\rho =\rho _{0}}, \\
J_{\mathrm{sym,4}} &=&27\rho _{0}^{3}\frac{d^{3}E_{\mathrm{sym,4}}(\rho )}
{d\rho ^{3}}|_{\rho =\rho _{0}}, \\
I_{\mathrm{sym,4}} &=&81\rho
_{0}^{4}\frac{d^{4}E_{\mathrm{sym,4}}(\rho )}{d\rho ^{4}}|_{\rho
=\rho _{0}}.
\end{eqnarray}%
\bigskip

In above Taylor expansions, we have kept all terms up to $4$th-order
in $\delta $ or $\chi $. The $14$ coefficients, namely, $E_{0}(\rho
_{0})$, $K_{0}$, $J_{0}$, $I_{0}$, $E_{\mathrm{sym}}(\rho _{0})$,
$L$, $K_{\mathrm{sym}}$, $J_{\mathrm{sym}}$, $I_{\mathrm{sym}}$,
$E_{\mathrm{sym,4}}(\rho _{0})$, $L_{\mathrm{sym,4}}$,
$K_{\mathrm{sym,4}}$, $J_{\mathrm{sym,4}}$ and $I_{\mathrm{sym,4}}$,
are well-defined, and they characterize the EOS of an asymmetric
nuclear matter and its density dependence at the normal nuclear
density $\rho _{0}$. Among these parameters, $E_{0}(\rho _{0})$,
$K_{0}$, $E_{\mathrm{sym}}(\rho _{0})$, $L$, and $K_{\mathrm{sym}}$
have been extensively studied in the literature and significant
progress has been made over past few decades. Based on these
well-defined quantities, we investigate in the present paper to what
extend they can provide reliable information on the isospin
dependence of the saturation properties of asymmetric nuclear matter
as well as its properties at both high and low densities.

\subsection{Saturation density}

One of the basic quantities for describing an asymmetric nuclear
matter is its saturation density $\rho _{\mathrm{sat}}(\delta )$
which is generally a function of the isospin asymmetry $\delta $.
According to the definition of the saturation density $\rho
_{\mathrm{sat}}(\delta )$ of asymmetric nuclear matter, i.e.,
\begin{equation}
\frac{\partial E(\rho ,\delta )}{\partial \rho }|_{\rho =\rho
_{\mathrm{sat}}(\delta )}=0,  \label{rhoSatDef}
\end{equation}%
it can be shown that up to $4$th-order in $\delta $, the saturation
density can be expressed as (see Appendix \ref{AppendA})
\begin{eqnarray}
\rho _{\mathrm{sat}}(\delta ) &=&\left[1-\frac{3L}{K_{0}}\delta ^{2}
+\left(\frac{3K_{\mathrm{sym}}L}{K_{0}^{2}}-\frac{3L_{\mathrm{sym,4}}}
{K_{0}}\right.\right.\notag\\
&&\left.\left.-\frac{3J_{0}L^{2}}{2K_{0}^{3}}\right)\delta
^{4}+O(\delta ^{6})\right]\rho _{0}.  \label{rhosat0}
\end{eqnarray}%
Therefore, $\rho _{\mathrm{sat}}(\delta )$ can be written as
\begin{equation}
\rho _{\mathrm{sat}}(\delta )=\rho _{0}+\rho _{\mathrm{sat,2}}\delta
^{2}+\rho _{\mathrm{sat,4}}\delta ^{4}+O(\delta ^{6})  \label{rhosat}
\end{equation}%
with%
\begin{eqnarray}
\rho _{\mathrm{sat,2}} &=&-\frac{3L}{K_{0}}\rho _{0}  \label{rhosat2} \\
\rho _{\mathrm{sat,4}} &=&\left(\frac{3K_{\mathrm{sym}}L}{K_{0}^{2}}
-\frac{3L_{\mathrm{sym,4}}}{K_{0}}-\frac{3J_{0}L^{2}}{2K_{0}^{3}}\right)\rho
_{0} \label{rhosat4}
\end{eqnarray}%
which reflects the shift of the saturation density of asymmetric
nuclear matter relative to that of symmetric nuclear matter due to
the finite isospin asymmetry.

\subsection{Binding energy at saturation density}

Another basic quantity of asymmetric nuclear matter is the binding
energy per nucleon at saturation density, i.e.,
$E_{\mathrm{sat}}(\delta )$, and it is generally a function of the
isospin asymmetry $\delta $. According to Eq. (\ref{EOSANM}) and Eq.
(\ref{rhosat}), $E_{\mathrm{sat}}(\delta )$ can be expressed up to
$4$th-order in $\delta $ as
\begin{eqnarray}
&&E_{\mathrm{sat}}(\delta ) =E_{0}(\rho _{0})+\frac{K_{0}}{2\times
9}\left(
\frac{\rho _{\mathrm{sat}}-\rho _{0}}{\rho _{0}}\right) ^{2}\notag \\
&&+\left[ E_{\mathrm{sym}}(\rho _{0})+\frac{L}{3}\left( \frac{\rho _{\mathrm{%
sat}}-\rho _{0}}{\rho _{0}}\right) \right] \delta ^{2}+E_{\mathrm{sym,4}%
}(\rho _{0})\delta ^{4} \notag\\
&=&E_{0}(\rho _{0})+\frac{1}{2\times 9}\frac{9L^{2}}{K_{0}}\delta ^{4}+E_{%
\mathrm{sym}}(\rho _{0})\delta ^{2} \notag\\
&&-\frac{L}{3}\frac{3L}{K_{0}}\delta ^{4}+E_{\mathrm{sym,4}}(\rho
_{0})\delta ^{4}+O(\delta ^{6}) \notag\\
&=&E_{0}(\rho _{0})+E_{\mathrm{sym}}(\rho _{0})\delta ^{2}+\left( E_{\mathrm{%
sym,4}}(\rho _{0})-\frac{L^{2}}{2K_{0}}\right) \delta ^{4} \notag\\
&&+O(\delta ^{6}),
\end{eqnarray}%
Therefore, $E_{\mathrm{sat}}(\delta )$ can be written as%
\begin{equation}
E_{\mathrm{sat}}(\delta )=E_{0}(\rho _{0})+E_{\mathrm{sat,2}}\delta ^{2}+E_{%
\mathrm{sat,4}}\delta ^{4}+O(\delta ^{6}),  \label{Esat}
\end{equation}%
with
\begin{eqnarray}
E_{\mathrm{sat,2}} &=&E_{\mathrm{sym}}(\rho _{0})  \label{Esat2} \\
E_{\mathrm{sat,4}} &=&E_{\mathrm{sym,4}}(\rho
_{0})-\frac{L^{2}}{2K_{0}}. \label{Esat4}
\end{eqnarray}%
The binding energy of asymmetric nuclear matter at saturation
density is thus shifted from that of symmetry nuclear matter as a
result of the shift of the saturation density when the isospin
asymmetry is finite.

\subsection{Incompressibility at saturation density}

The incompressibility of asymmetric nuclear matter is an important
quantity to describe its properties. It depends on the density and
isospin asymmetry of the asymmetric nuclear matter and is
characterized by the incompressibility coefficient defined as
\begin{eqnarray}
K(\rho ,\delta ) &=&9\frac{\partial P(\rho ,\delta )}{\partial \rho
} \notag \\
&=&18\rho \frac{\partial E(\rho ,\delta )}{\partial \rho }+9\rho ^{2}\frac{%
\partial ^{2}E(\rho ,\delta )}{\partial \rho ^{2}}  \notag \\
&=&18\frac{P(\rho ,\delta )}{\rho }+9\rho ^{2}\frac{\partial ^{2}E(\rho
,\delta )}{\partial \rho ^{2}},
\end{eqnarray}%
where $P(\rho ,\delta )$ is the pressure of asymmetric nuclear
matter and it can be expressed as
\begin{equation}
P(\rho ,\delta )=\rho ^{2}\frac{\partial E(\rho ,\delta )}{\partial \rho }.
\end{equation}%
Conventionally, the incompressibility coefficient is defined at the
saturation density where $P(\rho ,\delta )=0$. It is also called the
isobaric incompressibility coefficient \cite{Pra85}, and is given by
\begin{equation}
K_{\mathrm{sat}}(\delta )=9\rho _{\mathrm{sat}}^{2}\frac{\partial
^{2}E(\rho ,\delta )}{\partial \rho ^{2}}|_{\rho =\rho
_{\mathrm{sat}}}. \label{KsatDef}
\end{equation}%
The isobaric incompressibility coefficient $K_{\mathrm{sat}}(\delta
)$ thus only depends on the isospin asymmetry $\delta $. One can
show (see Appendix \ref{AppendA}) that up to $4$th-order in $\delta
$, the isobaric incompressibility coefficient
$K_{\mathrm{sat}}(\delta )$ can be expressed as

\begin{eqnarray}
&&K_{\mathrm{sat}}(\delta)=K_{0}+(K_{\mathrm{sym}}-6L-\frac{J_{0}}{K_{0}}L)
\delta ^{2}  \notag \\
&&+(K_{\mathrm{sym,4}}-6L_{\mathrm{sym,4}}-\frac{J_{0}L_{\mathrm{sym,4}}}{%
K_{0}}+\frac{9L^{2}}{K_{0}}-\frac{J_{\mathrm{sym}}L}{K_{0}}  \notag \\
&&+\frac{I_{0}L^{2}}{2K_{0}^{2}}+\frac{J_{0}K_{\mathrm{sym}}L}{K_{0}^{2}}+%
\frac{3J_{0}L^{2}}{K_{0}^{2}}-\frac{J_{0}^{2}L^{2}}{2K_{0}^{3}})
\delta^4\notag\\
&&+O(\delta ^{6}),  \label{Ksat0}
\end{eqnarray}%
which can be further written as%
\begin{equation}
K_{\mathrm{sat}}(\delta )=K_{0}+K_{\mathrm{sat,2}}\delta ^{2}+K_{\mathrm{%
sat,4}}\delta ^{4}+O(\delta ^{6}) \label{Ksat}
\end{equation}%
with%
\begin{eqnarray}
K_{\mathrm{sat,2}} &=&K_{\mathrm{sym}}-6L-\frac{J_{0}}{K_{0}}L,
\label{Ksat2} \\
K_{\mathrm{sat,4}} &=&K_{\mathrm{sym,4}}-6L_{\mathrm{sym,4}}-\frac{J_{0}L_{%
\mathrm{sym,4}}}{K_{0}}+\frac{9L^{2}}{K_{0}}-\frac{J_{\mathrm{sym}}L}{K_{0}}
\notag \\
&&+\frac{I_{0}L^{2}}{2K_{0}^{2}}+\frac{J_{0}K_{\mathrm{sym}}L}{K_{0}^{2}}+%
\frac{3J_{0}L^{2}}{K_{0}^{2}}-\frac{J_{0}^{2}L^{2}}{2K_{0}^{3}}.
\label{Ksat4}
\end{eqnarray}%
The coefficient $K_{\mathrm{sat,2}}$ and $K_{\mathrm{sat,4}}$ reflect the
isospin dependence of the isobaric incompressibility of asymmetric nuclear
matter.

If we use the parabolic approximation for the EOS of symmetric
nuclear matter, i.e.,\ Eq. (\ref{E0para}), then the
$K_{\mathrm{sat,2}}$ parameter is reduced to
\begin{equation}
K_{\mathrm{asy}}=K_{\mathrm{sym}}-6L,  \label{Kasy}
\end{equation}%
and this expression has been extensively used in the literature to
characterize the isospin dependence of the incompressibility of
asymmetric nuclear matter
\cite{Lop88,Bar02,Bar05,Che05a,Dan09,Cen09}. Obviously, we have

\begin{equation}
K_{\mathrm{sat,2}}=K_{\mathrm{asy}}-\frac{J_{0}}{K_{0}}L,  \label{Ksat2Kasy}
\end{equation}%
and the coefficient $K_{\mathrm{asy}}$ thus could be a good
approximation to $K_{\mathrm{sat,2}}$ if the $3$rd-order derivative
of the EOS of symmetric nuclear matter with respect to density,
i.e., $J_{0}$, is negligible or the magnitude of the slope parameter
$L$ for the symmetry energy is very small. It should be noted that
the higher-order derivatives of the EOS of symmetric nuclear matter
with respect to density in Eq. (\ref{E0}), such as $I_{0}$, do not
contribute to\ $K_{\mathrm{sat,2}}$. In the following, we will check
how the $J_{0}$ term affects the value of $K_{\mathrm{sat,2}}$.

As shown in Appendix \ref{AppendA}, the expressions Eq.
(\ref{rhosat}) for the saturation density, Eq. (\ref{Esat}) for the
binding energy and Eq. (\ref{Ksat}) for the isobaric
incompressibility coefficient are exact up to $4$th-order in $\delta
$. It is thus interesting to see that with precision up to
$4$th-order in $\delta $, we only need to know $11$ coefficients
$E_{0}(\rho _{0})$, $K_{0}$, $J_{0}$, $I_{0}$,
$E_{\mathrm{sym}}(\rho _{0})$, $L$, $K_{\mathrm{sym}}$,
$J_{\mathrm{sym}}$, $E_{\mathrm{sym,4}}(\rho _{0})$,
$L_{\mathrm{sym,4}}$, $K_{\mathrm{sym,4}}$ among the $14$
coefficients $E_{0}(\rho _{0})$, $K_{0}$, $J_{0}$, $I_{0}$,
$E_{\mathrm{sym}} (\rho _{0})$, $L$, $K_{\mathrm{sym}}$,
$J_{\mathrm{sym}}$, $I_{\mathrm{sym}} $, $E_{\mathrm{sym,4}}(\rho
_{0})$, $L_{\mathrm{sym,4}}$, $K_{\mathrm{sym,4}} $,
$J_{\mathrm{sym,4}}$ and $I_{\mathrm{sym,4}}$ which are defined at
the normal nuclear density $\rho _{0}$. The higher-order coefficient
$I_{\mathrm{sym}}$ in Eq. (\ref{EsymLKJI}) for the symmetry energy
as well as $J_{\mathrm{sym,4}}$ and $I_{\mathrm{sym,4}}$ in
Eq.(\ref{Esym4LKJI}) for the $4$th-order symmetry energy do not
affect the saturation density, the binding energy and the isobaric
incompressibility coefficient with precision up to $4$th-order in
$\delta $. Furthermore, the $6$ coefficients $E_{0}(\rho _{0})$,
$K_{0}$, $J_{0}$, $E_{\mathrm{sym}}(\rho _{0})$, $L$, and
$K_{\mathrm{sym}}$ determine completely the saturation density, the
binding energy and the isobaric incompressibility coefficient with
precision up to $2$nd-order in $\delta $.

\section{Theoretical models}

\label{models}

In this section, we introduce the theoretical models used in the
present work and also give some important expressions for
completeness. These models include the modified finite-range Gogny
effective interaction (MDI)~\cite{Das03}, the Hartree-Fock approach
based on Skyrme interactions, and a phenomenological modified
Skyrme-like (MSL) model. A very useful feature of these models is
that analytical expressions for many interesting physical quantities
in asymmetric nuclear matter at zero temperature can be obtained,
and this makes it physically transparent and very convenient to
check the higher-order effects on the properties of asymmetric
nuclear matter.

\subsection{Isospin- and momentum-dependent MDI interaction}

The isospin- and momentum-dependent MDI interaction is based on the
finite-range Gogny effective interaction~\cite{Das03}. In the MDI
interaction, the potential energy density $V(\rho ,\delta )$ of an
asymmetric nuclear matter at total density $\rho $ and isospin
asymmetry $\delta $ is expressed as follows~\cite{Das03,Che05a},
\begin{eqnarray}
V(\rho ,\delta ) &=&\frac{A_{u}\rho _{n}\rho _{p}}{\rho _{0}}+\frac{A_{l}}{%
2\rho _{0}}(\rho _{n}^{2}+\rho _{p}^{2})+\frac{B}{\sigma +1}\frac{\rho
^{\sigma +1}}{\rho _{0}^{\sigma }}  \notag \\
&\times &(1-x\delta ^{2})+\frac{1}{\rho _{0}}\sum_{\tau ,\tau ^{\prime
}}C_{\tau ,\tau ^{\prime }}  \notag \\
&\times &\int \int d^{3}pd^{3}p^{\prime }\frac{f_{\tau }(\vec{r},\vec{p}%
)f_{\tau ^{\prime }}(\vec{r},\vec{p}^{\prime })}{1+(\vec{p}-\vec{p}^{\prime
})^{2}/\Lambda ^{2}}.  \label{MDIV}
\end{eqnarray}%
In the above, $\tau =1/2$ ($-1/2$) for neutrons (protons); $\sigma
=4/3$; $f_{\tau }(\vec{r},\vec{p})$ is the phase-space distribution
function of neutrons or protons at coordinate $\vec{r}$ and momentum
$\vec{p}$. The parameters $A_{u}(x),A_{l}(x),B,C_{\tau ,\tau
},C_{\tau ,-\tau }$ and $\Lambda $ are obtained by fitting the
momentum-dependence of the single-nucleon potential $U(\rho ,\delta
,\vec{p},\tau )$ to that predicted by the Gogny Hartree-Fock and/or
the Brueckner-Hartree-Fock calculations, the saturation properties
of symmetric nuclear matter, and the symmetry energy at the normal
nuclear matter density $\rho _{0}$ \cite{Das03}. The parameters
$A_{u}(x)$ and $A_{l}(x)$ are given by
\begin{equation}
A_{u}(x)=-95.98-x\frac{2B}{\sigma
+1},~A_{l}(x)=-120.57+x\frac{2B}{\sigma +1}
\end{equation}%
in terms of the parameter $x$, which is introduced to mimic various
$E_{\mathrm{sym}}(\rho )$ predicted by different microscopic and/or
phenomenological many-body theories~\cite{Che05a}. By adjusting the
$x$ parameter, the $E_{\mathrm{sym}}(\rho )$ is varied without
changing any property of symmetric nuclear matter and the symmetry
energy at the saturation density $E_{\mathrm{sym}}(\rho _{0})$, as
the $x$-dependent $A_{u}(x)$ and $A_{l}(x)$ are adjusted
accordingly. Using the definition in Eq.~(\ref{Esym}), we have
$E_{\mathrm{sym}}(\rho _{0})=30.5 $ MeV at $\rho _{0}=0.16$
fm$^{-3}$ while its value is $31.6$ MeV within the parabolic
approximation of $E_{\mathrm{sym}}(\rho )\approx E(\rho ,\delta
=1)-E(\rho ,\delta =0)$. We note that the MDI interaction has been
extensively used in the transport model for studying isospin effects
in intermediate energy heavy-ion collisions induced by neutron-rich
nuclei
\cite{LiBA04a,Che04,Che05a,LiBA05c,LiBA05a,LiBA05b,LiBA06b,Yon06a,Yon06b,Yon07,CKLY07,LCK08},
the study of the thermal properties of asymmetric nuclear
matter~\cite{Xu07,Xu07b,LiBA07a}, and the transition density and
pressure in neutron star crust~\cite{Xu09a,Xu09b}. In particular,
the isospin diffusion data from NSCL/MSU have constrained the value
of $x$ to between $0$ and $-1$ for nuclear densities from about
$0.3\rho _{0}$ to $1.2\rho _{0}$ \cite{Che05a,LiBA05c}.

With $f_{\tau }(\vec{r},\vec{p})$ $=\frac{2}{h^{3}}\Theta
(p_{f}(\tau )-p)$ for nuclear matter at zero temperature, the
integrals in Eq.~(\ref{MDIV}) can be evaluated analytically. In
particular, we have \cite{Che07}
\begin{eqnarray}
&\int \int &d^{3}pd^{3}p^{\prime }\frac{f_{\tau }(\vec{r},\vec{p})f_{\tau
^{\prime }}(\vec{r},\vec{p}^{\prime })}{1+(\vec{p}-\vec{p}^{\prime
})^{2}/\Lambda ^{2}}  \notag \\
&=&\frac{1}{6}\left( \frac{4\pi }{h^{3}}\right) ^{2}\Lambda
^{2}\left\{p_{f}(\tau )p_{f}(\tau ^{\prime })[3(p_{f}^{2}(\tau
)+p_{f}^{2}(\tau ^{\prime}))-\Lambda ^{2}]\right.  \notag \\
&+&4\Lambda \lbrack (p_{f}^{3}(\tau )-p_{f}^{3}(\tau ^{\prime }))\arctan
\frac{p_{f}(\tau )-p_{f}(\tau ^{\prime })}{\Lambda }  \notag \\
&-&\left.(p_{f}^{3}(\tau )+p_{f}^{3}(\tau ^{\prime }))\arctan
\frac{p_{f}(\tau)+p_{f}(\tau ^{\prime })}{\Lambda }\right]  \notag \\
&+&\frac{1}{4}[\Lambda ^{4}+6\Lambda ^{2}(p_{f}^{2}(\tau )+p_{f}^{2}(\tau
^{\prime }))-3(p_{f}^{2}(\tau )-p_{f}^{2}(\tau ^{\prime }))^{2}]  \notag \\
&\times &\left.\ln \frac{(p_{f}(\tau )+p_{f}(\tau ^{\prime }))^{2}
+\Lambda ^{2}}{(p_{f}(\tau )-p_{f}(\tau ^{\prime }))^{2}+\Lambda
^{2}}\right\}.
\end{eqnarray}%
The kinetic energy per nucleon of cold asymmetric nuclear matter is
\begin{eqnarray}
E_{k}(\rho ,\delta ) &=&\frac{1}{\rho }\int d^{3}p\left( \frac{p^{2}}{2m}%
f_{n}(\vec{r},\vec{p})+\frac{p^{2}}{2m}f_{p}(\vec{r},\vec{p})\right)  \notag
\\
&=&\frac{4\pi }{5mh^{3}\rho }(p_{fn}^{5}+p_{fp}^{5}),
\end{eqnarray}%
where $p_{fn(p)}=\hbar (3\pi ^{2}\rho _{n(p)})^{1/3}$ is the Fermi
momentum of neutrons (protons). The total energy per nucleon of cold
asymmetric nuclear matter can be obtained as
\begin{equation}
E(\rho ,\delta )=\frac{V(\rho ,\delta )}{\rho }+E_{k}(\rho ,\delta ).
\label{EOSMDI}
\end{equation}%
By setting $\rho _{n}=\rho _{p}=\frac{\rho }{2}$ and
$p_{fn}=p_{fp}=p_{f}$, where $p_{f}=\hbar (3\pi ^{2}\rho/2)^{1/3}$
is the fermi momentum of symmetric nuclear matter at density $\rho
$, we obtain following EOS for cold symmetric nuclear matter:
\begin{widetext}
\begin{eqnarray}
E_{0}(\rho ) &=&\frac{8\pi }{5mh^{3}\rho }p_{f}^{5}+\frac{\rho }{4\rho _{0}}%
(A_{l}+A_{u})+\frac{B}{\sigma +1}\left( \frac{\rho }{\rho _{0}}\right)
^{\sigma }  \notag \\
&+&\frac{1}{3\rho _{0}\rho }(C_{l}+C_{u})\left( \frac{4\pi }{h^{3}}\right)
^{2}\Lambda ^{2}  \notag \\
&\times &\{p_{f}^{2}(6p_{f}^{2}-\Lambda ^{2})-8\Lambda p_{f}^{3}\arctan
\frac{2p_{f}}{\Lambda }  \notag \\
&+&\left.\frac{1}{4}[\Lambda ^{4}+12\Lambda ^{2}p_{f}^{2}]\ln \frac{%
4p_{f}^{2}+\Lambda ^{2}}{\Lambda ^{2}}\right\}.
\end{eqnarray}%
Furthermore, from Eq.~(\ref{EOSMDI}) the symmetry energy can be
expressed as
\begin{eqnarray}
E_{\mathrm{sym}}(\rho ) &=&\frac{1}{2}\left( \frac{\partial ^{2}E}{\partial
\delta ^{2}}\right) _{\delta =0}  \notag \\
&=&\frac{8\pi }{9mh^{3}\rho }p_{f}^{5}+\frac{\rho }{4\rho
_{0}}(A_{l}-A_{u})- \frac{Bx}{\sigma +1}\left( \frac{\rho }{\rho
_{0}}\right) ^{\sigma }  \notag\\
&+&\frac{C_{l}}{9\rho _{0}\rho }\left( \frac{4\pi }{h^{3}}\right)
^{2}\Lambda ^{2}\left[ 4p_{f}^{4}-\Lambda ^{2}p_{f}^{2}\ln \frac{%
4p_{f}^{2}+\Lambda ^{2}}{\Lambda ^{2}}\right]  \notag \\
&+&\frac{C_{u}}{9\rho _{0}\rho }\left( \frac{4\pi }{h^{3}}\right)
^{2}\Lambda ^{2}\left[ 4p_{f}^{4}-p_{f}^{2}(4p_{f}^{2}+\Lambda
^{2})\ln \frac{4p_{f}^{2}+\Lambda ^{2}}{\Lambda ^{2}}\right] ,
\label{EsymMDI}
\end{eqnarray}%
and the $4$th-order nuclear symmetry energy can be written as
\begin{eqnarray}
E_{\mathrm{sym,4}}(\rho ) &=&\frac{1}{4!}\left( \frac{\partial ^{4}E}{%
\partial \delta ^{4}}\right) _{\delta =0}  \notag \\
&=&\frac{8\pi }{3^{5}mh^{3}\rho }p_{f}^{5}  \notag \\
&-&\frac{C_{l}}{3^{5}\rho _{0}\rho }\left( \frac{4\pi }{h^{3}}\right)
^{2}\Lambda ^{2}\left[ 7\Lambda ^{2}p_{f}^{2}\ln \frac{4p_{f}^{2}+\Lambda
^{2}}{\Lambda ^{2}}-\frac{4(7\Lambda ^{4}p_{f}^{4}+42\Lambda
^{2}p_{f}^{6}+40p_{f}^{8}}{(4p_{f}^{2}+\Lambda ^{2})^{2}}\right]  \notag \\
&-&\frac{C_{u}}{3^{5}\rho _{0}\rho }\left( \frac{4\pi }{h^{3}}\right)
^{2}\Lambda ^{2}\left[ (7\Lambda ^{2}p_{f}^{2}+16p_{f}^{4})\ln \frac{%
4p_{f}^{2}+\Lambda ^{2}}{\Lambda ^{2}}-28p_{f}^{4}-\frac{8p_{f}^{6}}{\Lambda
^{2}}\right] .  \label{Esym4MDI}
\end{eqnarray}%

\end{widetext}

\subsection{Skyrme-Hartree-Fock approach}

In the standard Skyrme Hartree-Fock model \cite
{Vau72,Bra85,Fri86,Cha97,Cha98,Bro98,Che99b,Bro00,Sto03,Sto07}, the
interaction is taken to have a zero-range, density- and
momentum-dependent form with the interaction parameters determined
from fitting the binding energies and charge radii of a large number
of nuclei in the periodic table. In this approach, the EOS of
asymmetric nuclear matter can be expressed
as~\cite{Cha97,Cha98,Sto03,Sto07}
\begin{eqnarray}
E(\rho ,\delta ) &=&\frac{3\hbar ^{2}}{10m}\left( \frac{3\pi ^{2}}{2}\right)
^{2/3}\rho ^{\frac{2}{3}}F_{5/3}  \notag \\
&+&\frac{1}{8}t_{0}\rho \lbrack 2(x_{0}+2)-(2x_{0}+1)F_{2}]  \notag \\
&+&\frac{1}{48}t_{3}\rho ^{\sigma +1}[2(x_{3}+2)-(2x_{3}+1)F_{2}]  \notag \\
&+&\frac{3}{40}\left( \frac{3\pi ^{2}}{2}\right) ^{2/3}\rho ^{\frac{5}{3}}
\notag \\
&\times &\{[t_{1}(x_{1}+2)+t_{2}(x_{2}+2)]F_{5/3}  \notag \\
&+&\frac{1}{2}[t_{2}(2x_{2}+1)-t_{1}(2x_{1}+1)]F_{8/3}\},
\end{eqnarray}%
where
\begin{equation}
F_{m}(\delta )=\frac{1}{2}[(1+\delta )^{m}+(1-\delta )^{m}].  \notag
\end{equation}%
The EOS of symmetric nuclear matter can thus be written as

\begin{eqnarray}
E_{0}(\rho ) &=&\frac{3\hbar ^{2}}{10m}\left( \frac{3\pi ^{2}}{2}\right)
^{2/3}\rho ^{\frac{2}{3}}+\frac{3}{8}t_{0}\rho  \notag \\
&+&\frac{3}{80}\Theta _{s}\left( \frac{3\pi ^{2}}{2}\right) ^{2/3}\rho ^{%
\frac{5}{3}}+\frac{1}{16}t_{3}\rho ^{\sigma +1},
\end{eqnarray}%
with $\Theta _{s}=3t_{1}+(5+4x_{2})t_{2}$. Furthermore, the symmetry
energy can be obtained as

\begin{eqnarray}
E_{\text{\textrm{sym}}}(\rho ) &=&\frac{1}{2}\left( \frac{\partial ^{2}E}{%
\partial \delta ^{2}}\right) _{\delta =0}  \notag \\
&=&\frac{\hbar ^{2}}{6m}\left( \frac{3\pi ^{2}}{2}\right) ^{2/3}\rho ^{\frac{%
2}{3}}-\frac{1}{8}t_{0}(2x_{0}+1)\rho  \notag \\
&-&\frac{1}{24}\left( \frac{3\pi ^{2}}{2}\right) ^{2/3}
\Theta _{\rm sym}\rho ^{\frac{5}{3}}  \notag \\
&-&\frac{1}{48}t_{3}(2x_{3}+1)\rho ^{\sigma +1},
\end{eqnarray}%
with $\Theta _{\rm sym}=3t_{1}x_{1}-t_{2}(4+5x_{2})$. Similarly, the
$4$th-order nuclear symmetry energy can be written as
\begin{eqnarray}
E_{\mathrm{sym,4}}(\rho ) &=&\frac{1}{4!}\left( \frac{\partial ^{4}E}{%
\partial \delta ^{4}}\right) _{\delta =0}  \notag \\
&=&\frac{\hbar ^{2}}{162m}\left( \frac{3\pi ^{2}}{2}\right) ^{2/3}\rho ^{%
\frac{2}{3}}  \notag \\
&&+\frac{1}{648}\left( \frac{3\pi ^{2}}{2}\right) ^{2/3}\Theta _{\rm
sym,4}\rho ^{\frac{5}{3}},  \notag
\end{eqnarray}%
with $\Theta_{\rm sym,4}=t_{1}(1-x_{1})+3t_{2}(1+x_{2})$.

\subsection{A phenomenological modified Skyrme-like (MSL) model}

Following the energy density functional obtained from the
Hartree-Fock approach with the zero-range, density- and
momentum-dependent form with the Skyrme interaction, the binding
energy per nucleon of a cold asymmetric nuclear matter at total
density $\rho $ and isospin asymmetry $\delta $ in the modified
Skyrme-like (MSL) model is parameterized as
\begin{eqnarray}
E_{\text{\textrm{MSL}}}(\rho ,\delta ) &=&\frac{\eta }{\rho }\left( \frac{%
\hbar ^{2}}{2m_{n}^{\ast }}\rho _{n}^{5/3}+\frac{\hbar ^{2}}{2m_{p}^{\ast }}%
\rho _{p}^{5/3}\right)  \notag \\
&&+\frac{\alpha }{2}\frac{\rho }{\rho _{0}}+\frac{\beta }{\sigma +1}\frac{%
\rho ^{\sigma }}{{\rho _{0}}^{\sigma }}+E_{\text{\textrm{sym}}}^{\rm
loc} ({\rho })\delta ^{2},  \label{EOSMID}
\end{eqnarray}%
where $\eta =\frac{3}{5}\left( 3\pi ^{2}\right) ^{2/3}$; $\alpha $,
$\beta $ and $\sigma $ are parameters; and
$E_{\text{\textrm{sym}}}^{\rm loc}({\rho })$ represents the local
density-dependent part of the symmetry energy. The effective neutron
and proton masses $m_{n}^{\ast }$ and $m_{p}^{\ast }$ are assumed to
have forms similar to the standard SHF results
\cite{Cha97,Cha98,Sto03,Sto07}, i.e.,

\begin{eqnarray}
\frac{\hbar ^{2}}{2m_{n}^{\ast }} &=&\frac{\hbar ^{2}}{2m}+\rho \left( C_{%
\mathrm{eff}}+D_{\mathrm{eff}}\delta \right) \\
\frac{\hbar ^{2}}{2m_{p}^{\ast }} &=&\frac{\hbar ^{2}}{2m}+\rho \left( C_{%
\mathrm{eff}}-D_{\mathrm{eff}}\delta \right),
\end{eqnarray}%
where $C_{\mathrm{eff}}$ and $D_{\mathrm{eff}}$ are constants. This
implies that the single-nucleon potential depends quadratically on
the nucleon momentum as in the standard SHF approach. In terms of
the isoscalar effective mass $m_{s}^{\ast }$ and the isovector
effective mass $m_{v}^{\ast }$ given by
\begin{eqnarray}
\frac{\hbar ^{2}}{2m_{s}^{\ast }} &=&\frac{\hbar ^{2}}{2m}+\rho C_{\mathrm{%
eff}} \\
\frac{\hbar ^{2}}{2m_{v}^{\ast }} &=&\frac{\hbar ^{2}}{2m}+\rho \left( C_{%
\mathrm{eff}}-D_{\mathrm{eff}}\right),
\end{eqnarray}%
the nucleon effective mass can be written as \cite{Pea01}
\begin{equation}
\frac{\hbar ^{2}}{2m_{q}^{\ast }}=\frac{2\rho _{q}}{\rho }\frac{\hbar ^{2}}{%
2m_{s}^{\ast }}+\left( 1-\frac{2\rho _{q}}{\rho }\right) \frac{\hbar ^{2}}{%
2m_{v}^{\ast }},\quad q=n,p.
\end{equation}%
We note that the isovector effective mass $m_{v}^{\ast }$
corresponds to the proton (neutron) effective mass in pure neutron
(proton) matter. Also, we can easily obtain the following relation
\begin{equation}
\frac{\hbar ^{2}}{2m_{n}^{\ast }}-\frac{\hbar ^{2}}{2m_{p}^{\ast }}=2\delta
\left( \frac{\hbar ^{2}}{2m_{s}^{\ast }}-\frac{\hbar ^{2}}{2m_{v}^{\ast }}%
\right) .  \label{EffNPdiff}
\end{equation}%
Experimentally, the isoscalar effective mass $m_{s}^{\ast }$ and the
isovector effective mass $m_{v}^{\ast }$ at the normal nuclear
density $\rho
_{0}$ have been constrained to be $m_{s,0}^{\ast }\approx 0.8m$ and $%
m_{v,0}^{\ast }\approx 0.7m$, respectively~\cite%
{Boh79,Kri80,Cha97,Cha98,Rei99,Far01,Lun03,Mar07,Che07}. With these
constrained values for the isoscalar and isovector effective masses,
Eq.~(\ref{EffNPdiff}) gives a larger neutron effective mass than the
proton effective mass in a neutron-rich matter, which is consistent
with experimental data on the isospin dependence of the nucleon
optical potential and also recent microscopic and phenomenological
many-body theory predictions \cite{LiBA04c,Fuc05,LCK08}.

The EOS of symmetric nuclear matter in the MSL model is then given
by
\begin{eqnarray}
E_{0}(\rho ) &=&E_{\rm kin}^{0}\left( \frac{{\rho }}{{\rho
_{0}}}\right)
^{2/3}+C\left( \frac{{\rho }}{{\rho _{0}}}\right) ^{5/3}  \notag \\
&&+\frac{\alpha }{2}\frac{\rho }{\rho _{0}}+\frac{\beta }{\sigma
+1}\left( \frac{\rho }{{\rho _{0}}}\right) ^{\sigma },
\label{E0MID}
\end{eqnarray}%
where the first term represents the kinetic energy contribution with
$E_{\rm kin}^{0}=\frac{3\hbar ^{2}}{10m}\left( \frac{3\pi
^{2}}{2}\right) ^{2/3}\rho _{0}^{2/3}$ and the second term is due to
the nucleon effective mass with the coefficient $C$ being a constant
determined by the isoscalar effective mass $m_{s,0}^{\ast }$ as
\begin{equation}
C=\frac{m-m_{s,0}^{\ast }}{m_{s,0}^{\ast }}E_{\rm kin}^{0}.
\end{equation}%

The parameters $\alpha $, $\beta $ and $\sigma $ in the MSL model
are determined by the binding energy per nucleon $E_{0}(\rho _{0})$
and the incompressibility $K_{0}$ of cold symmetric nuclear matter
at the saturation density $\rho _{0}$, and they can be expressed as
\begin{eqnarray}
\alpha &=&-\frac{4}{3}E_{\rm kin}^{0}-\frac{10}{3}C-\frac{2}{3}%
(E_{\rm kin}^{0}-3E_{0}(\rho _{0})-2C)  \notag \\
&&\times \frac{K_{0}+2E_{\rm kin}^{0}-10C}{K_{0}+9E_{0}(\rho _{0})
-E_{\rm kin}^{0}-4C} \\
\beta &=&(\frac{E_{\rm kin}^{0}}{3}-E_{0}(\rho _{0})-\frac{2}{3}C)  \notag \\
&&\times \frac{K_{0}-9E_{0}(\rho _{0})+5E_{\rm
kin}^{0}-16C}{K_{0}+9E_{0}(\rho_{0})-E_{\rm kin}^{0}-4C} \\
\sigma &=&\frac{K_{0}+2E_{\rm kin}^{0}-10C}{3E_{\rm
kin}^{0}-9E_{0}(\rho _{0})-6C}.
\end{eqnarray}%
In particular, for $E_{0}(\rho _{0})=-16$ MeV, $m_{s,0}^{\ast
}=0.8m$ and $\rho _{0}=0.16$ fm$^{-3}$, we have
\begin{eqnarray}
C &=&5.53\text{ (MeV)} \\
\alpha &=&-47.90-39.37\frac{K_{0}-11.05}{K_{0}-188.21}\text{ (MeV)} \\
\beta &=&19.68\frac{K_{0}+166.11}{K_{0}-188.21}\text{ (MeV)} \\
\sigma &=&\frac{K_{0}-11.05}{177.16},
\end{eqnarray}%
where the units of $K_{0}$ is MeV.

The symmetry energy in the MSL model can be expressed as
\begin{equation}
E_{\text{\textrm{sym}}}(\rho )=E_{\text{\textrm{sym}}}^{\rm
kin}({\rho _{0}}) \left( \frac{{\rho }}{{\rho _{0}}}\right)
^{2/3}+D\left( \frac{{\rho }}{{\rho _{0}}}\right)
^{5/3}+E_{\text{\textrm{sym}}}^{\rm loc}({\rho }), \label{EsymMID}
\end{equation}%
where the first term is the kinetic energy contribution with
$E_{\text{\textrm{sym}}}^{kin}({\rho _{0}})=\frac{\hbar
^{2}}{6m}\left( \frac{3\pi ^{2}}{2}{\rho _{0}}\right) ^{2/3}$ and
the second term is due to the contribution of the nucleon effective
mass with the coefficient $D$ determined by both $m_{s,0}^{\ast }$
and $m_{v,0}^{\ast }$ as
\begin{equation}
D=\frac{5}{9}E_{\rm kin}^{0}\left( 4\frac{m}{m_{s,0}^{\ast }}-3\frac{m}{%
m_{v,0}^{\ast }}-1\right) .
\end{equation}%
With $m_{s,0}^{\ast }=0.8m$ and $m_{v,0}^{\ast }=0.7m$ at $\rho
_{0}=0.16$ fm$^{-3}$,  we have $D=-3.51$ MeV. In
Eq.~(\ref{EsymMID}), similarly to the momentum-independent MID model
\cite{Che09}, the local density-dependent part
$E_{\text{\textrm{sym}}}^{\rm loc}({\rho })$ is
parameterized as%
\begin{equation}
E_{\text{\textrm{sym}}}^{\rm loc}({\rho
})=(1-y)E_{\text{\textrm{sym}}}^{\rm loc} ({\rho _{0}})\frac{{\rho
}}{{\rho _{0}}}+yE_{\text{\textrm{sym}}}^{\rm loc}({\rho
_{0}})\left( \frac{{\rho }}{{\rho _{0}}}\right) ^{\gamma
_{\mathrm{sym}}} \label{EsymDenMID}
\end{equation}%
with the constant $E_{\text{\textrm{sym}}}^{\rm loc}({\rho _{0}})$
determined by
\begin{equation}
E_{\text{\textrm{sym}}}^{\rm loc}({\rho
_{0}})=E_{\text{\textrm{sym}}}({\rho
_{0}})-E_{\text{\textrm{sym}}}^{\rm kin}({\rho _{0}})-D.
\end{equation}%
Obviously, we have $E_{\text{\textrm{sym}}}^{\rm loc}({\rho
_{0}})=21.2$ MeV following $E_{\text{\textrm{sym}}}({\rho _{0}})=30$
MeV and $E_{\text{\textrm{sym}}}^{kin}({\rho _{0}})=12.3$ MeV at
$\rho _{0}=0.16$ fm$^{-3}$. The default value for the $\gamma
_{\mathrm{sym}}$ parameter is taken to be $4/3$ in the MSL model
following the $E_{\text{\textrm{sym}}}({\rho })$ in the MDI
interaction, namely, Eq.~(\ref{EsymMDI}) (we will see how the
$\gamma _{\mathrm{sym}}$ parameter affects the symmetry energy in
the following). In particular, similarly to the $x$ parameter in the
MDI interaction, the dimensionless parameter $y$ in the MSL model is
introduced to mimic various $E_{\mathrm{sym}}(\rho )$ predicted by
different microscopic and/or phenomenological many-body theories for
a fixed $\gamma _{\mathrm{sym}}$ parameter. As we will show later,
for $\gamma _{\mathrm{sym}}=4/3$, adjusting the $y$ value can nicely
reproduce the $E_{\mathrm{sym}}(\rho )$ in the MDI interaction with
$x=-1$, $0$, and $1$. Therefore, the symmetry energy density
functional constructed in the MSL model is very flexible and can
mimic very different density behaviors by varying only one
parameter.

In the MSL model, similarly to the SHF approach, the $4$th-order and
higher-order nuclear symmetry energies only include contributions
from the kinetic energy and the nucleon effective mass while the
local density-dependent part of higher-order symmetry energies are
neglected in Eq.~(\ref{EOSMID}). In particular, the $4$th-order
nuclear symmetry energy in the MSL model can be shown to be
\begin{eqnarray}
E_{\mathrm{sym,4}}(\rho ) &=&\frac{1}{4!}\left( \frac{\partial ^{4}E}{%
\partial \delta ^{4}}\right) _{\delta =0}  \notag \\
&=&\frac{5}{243}E_{\rm kin}^{0}\left( \frac{{\rho }}{{\rho
_{0}}}\right) ^{2/3}+\frac{5}{243}C\left( \frac{{\rho }}{{\rho
_{0}}}\right) ^{5/3}, \label{Esym4MID}
\end{eqnarray}%
where the first term is the kinetic energy contribution while the
second term is due to the contribution from the nucleon effective
mass.

The MSL model is thus an extension of the momentum-independent MID
model \cite{Che09} by including the effects of the nucleon effective
mass. It provides a simple phenomenological parametrization of the
EOS of asymmetric nuclear matter and is thus a convenient and
transparent way to investigate the possible correlations among
higher-order and lower-order characteristic parameters of asymmetric
nuclear matter. In the MSL model, we have totally $8$ free
parameters, i.e., $C$, $D$, $\alpha $, $\beta $, $\sigma $,
$E_{\text{\textrm{sym}}}({\rho _{0}})$, $y$, and $\gamma
_{\mathrm{sym}}$ which can be determined by empirical information on
the EOS of symmetric nuclear matter, the nucleon effective mass and
the density dependence of symmetry energy. In particular, the
parameters $C$ and $D$ (or equivalently $C_{\mathrm{eff}}$ and
$D_{\mathrm{eff}}$) are determined by the isoscalar effective masses
$m_{s}^{\ast }$ and the isovector effective mass $m_{v}^{\ast }$ at
the normal nuclear density $\rho _{0}$, i.e., $m_{s,0}^{\ast }$ and
$m_{v,0}^{\ast }$. The parameters $\alpha $, $\beta $ and $\sigma $
are determined by $E_{0}(\rho _{0})$, $K_{0} $ and $\rho _{0}$ while
the parameters $E_{\text{\textrm{sym}}}({\rho _{0}})$, $y$, and
$\gamma _{\mathrm{sym}}$ are introduced to mimic the density
dependence of different symmetry energies predicted by microscopic
and/or phenomenological many-body theories. As a default in the MSL
model, we use $m_{s,0}^{\ast }=0.8m$, $m_{v,0}^{\ast }=0.7m$, $\rho
_{0}=0.16$ fm$^{-3} $, $E_{0}(\rho _{0})=-16$ MeV, $K_{0}=240$ MeV,
$E_{\text{\textrm{sym}}}({\rho _{0}})=30$ MeV, and $\gamma
_{\mathrm{sym}}=4/3$ and vary the parameter $y$ to describe
different symmetry energies.

\section{Results and discussions}

\label{results}

\subsection{Characteristic parameters at normal nuclear density and EOS of
asymmetric nuclear matter}

\begin{widetext}

\begin{table}[tbp]
\caption{{\protect\small The saturation density $\rho _{0}$ and the
characteristic parameters $E_{0}(\rho _{0})$ (MeV), $K_{0}$ (MeV),
$J_{0}$ (MeV), $I_{0}$ (MeV), $K_{\mathrm{asy}}$ (MeV),
$K_{\mathrm{sat,2}}$ (MeV) and $K_{\mathrm{sat,2}}$ (MeV) at
saturation density for the MDI interaction with $x=-1, 0, 1$ and the
SHF predictions with $63$ standard Skyrme interactions. The small
differences from Table I of Ref.~\cite{Xu09b} for some Skyrme
interactions are due to the use of $0.17$ ($0.33$) as an
approximation of $1/6$ ($1/3$) for the $\sigma$ parameter in
Ref~\cite{Xu09b}. The results shown here are thus more accurate than
those in Ref~\cite{Xu09b}.}} \label{E0tab1}
\begin{tabular}{ccccccccc}
\hline\hline
Force & $\rho_0$ & $E_0(\rho_0)$ & \quad $K_0$ & \quad $J_0$ & \quad $I_0$ & $K_{asy}$ & $K_{sat,2}$ & $K_{sat,4}$ \quad \\
\hline
$$ MDI(1) & 0.160 & -16.2  &  212.5 & -447.6 & 2160.8 & -352.0 & -321.1 & -8.4 \\
$$ MDI(0) & 0.160 & -16.2  &  212.5 & -447.6 & 2160.8 & -443.1 & -316.3 & 52.9 \\
$$ MDI(-1) & 0.160 & -16.2  &  212.5 & -447.6 & 2160.8 & -534.3 & -311.4 & 214.4 \\
$$ Z & 0.159 & -16.0 &   330.3 & -65.0 & -348.2 & -359.6&-369.4&100.5 \\
$$ E$_{\sigma}$ & 0.163 & -16.0  &  248.6 & -352.4 & 1337.1  &  -236.6&-288.9&57.9 \\
$$ E &  0.159 & -16.1  &  333.5 & -63.0 & -356.3  &  -383.1 &-389.1&41.1\\
$$ Z$_{\sigma}$ & 0.163 & -15.9  &  233.3 & -369.0 & 1546.0  &  -225.1&-271.6&43.3 \\
$$ SVI & 0.143 & -15.8  &  363.6 & 153.5 & -1107.4 &   -427.3&-424.2&-4.4 \\
$$ Z$_{\sigma}^*$ & 0.162 & -16.0  &  234.9 & -369.2 & 1544.4  &  -305.5&-312.6&-7.1 \\
$$ SkSC4 &  0.161 & -15.9 &   234.7 & -380.8 & 1549.2 & -316.5&-320.1&-4.1 \\
$$ SI & 0.155 & -16.0  &  370.4 & 152.3 & -1129.5 & -469.2&-469.7&-14.2 \\
$$ BSk3  &  0.157 & -15.8 &   234.8 & -380.9 & 1529.8 & -347.6&-336.6&-9.2 \\
$$ BSk1  &  0.157 & -15.8 &   231.3 & -385.6 & 1588.7 & -325.0&-313&-8.4 \\
$$ SIII  &  0.145 & -15.9 &   355.4 & 101.4 & -903.0 & -453.2&-456.0&-20.2 \\
$$ BSk2  &  0.157 & -15.8 &   233.7 & -380.1 & 1542.4 & -344.8&-331.9&-9.7 \\
$$ MSk7  &  0.157 & -15.8 &   231.2 & -385.4 & 1587.3 & -331.0&-315.4&-8.6 \\
$$ BSk4  &  0.157 & -15.8 &   236.8 & -367.2 & 1466.9 & -341.2&-321.7&-11.4 \\
$$ BSk8  &  0.159 & -15.8 &   230.3 & -372.4 & 1578.2 & -310.0&-286.0&-16.8 \\
$$ BSk6  &  0.157 & -15.8 &   229.1 & -370.6 & 1571.3 & -316.3&-289.0&-16.8 \\
$$ BSk7  &  0.157 & -15.8 &   229.3 & -370.9 & 1572.8 & -317.3&-288.2&-17.0 \\
$$ SKP & 0.163 & -16.0  &  201.0 & -435.6& 2127.8 & -384.3 &-341.9&7.4\\
$$ BSk5  &  0.157 & -15.8 &   237.2 & -367.9 & 1470.3 & -368.8&-335.6&-7.5 \\
$$ SKXm  &  0.159 & -16.0  &  238.1 & -380.4 & 1542.2 & -435.3 &-384.0&10.3\\
$$ RATP  &  0.160  & -16.0  &  239.4 & -349.7 & 1451.5 & -385.5&-338.2&-6.6 \\
\hline\hline
\end{tabular}%
\end{table}
\begin{table}[tbp]
\caption{{\protect\small Continued with Table \ref{E0tab1}}}
\label{E0tab2}
\begin{tabular}{ccccccccc}
\hline\hline
Force & $\rho_0$ & $E_0(\rho_0)$ & \quad $K_0$ & \quad $J_0$ & \quad $I_0$ & $K_{asy}$ & $K_{sat,2}$ & $K_{sat,4}$ \quad \\
\hline
$$ SKX  & 0.155 & -16.1  &  271.1 & -297.4 & 904.0 & -451.2&-414.8&2.3 \\
$$ SKXce  &0.156  &-15.9&  268.2&    -294.6&   892.9&    -439.3 &-402.5&3.8\\
$$ BSk15  &0.159& -16.0&   241.6&    -363.1&   1457.0 &   -395.9&    -345.4&    -2.7\\
$$ BSk16  &0.159& -16.1&   241.7 &-363.6&  1459.9 &-396.6&  -344.2&   -1.7\\
$$ BSk10  &0.159& -15.9&   238.8 &-370.3&  1479.6 &-418.3&  -360.6&   12.7\\
$$ SGII  &0.158&  -15.6&   214.7 &-380.9&  1741.8 &-371.7&  -304.9  &17.8\\
$$ BSk12  &0.159& -15.9&   238.1 &-369.1&   1474.9 &-419.4&  -360.5&   14.0\\
$$ BSk11  &0.159   &-15.9&   238.1 &-369.2&  1475.3&-420.0&  -360.5 &14.6\\
$$ SLy10 &0.156&  -15.9&   229.7 &-358.3&  1559.6  &-374.7&  -314.2&   -24.6\\
$$ BSk13  &0.159& -15.9&   238.1 &-369.2&   1475.2 &-420.8&  -360.6&   15.5\\
$$ BSk9  &0.159&  -15.9 &       231.4 &-374.9&  1591.5  &-384.7&  -320.0&   -3.1\\
$$ BSk14  & 0.159&   -15.9&   239.3 &-358.7&  1434.8 &-415.5&  -349.7&   14.2\\
$$ SLy230a &0.160&    -16.0&   229.9 &-364.2&  1593.6 &-364.1&   -293.9&   -32.1\\
$$ SLy6  &0.159&  -15.9&   229.9 &-360.2&  1568.8 &-383.7&  -312.9&   -13.2\\
$$ SLy8  &0.160&  -16.0&   229.9 &-363.2&  1587.1 &-388.4&  -316.8&   -12.0\\
$$ SLy4  &0.160&  -16.0&   229.9 &-363.1&  1586.9 &-392.1&  -320.5 &-12.7\\
$$ SLy0  & 0.161& -16.0&   230.2 &-365.2&  1598.7 &-389.2&  -317.2&   -11.6\\
$$ SLy3  &0.160&  -16.0&   229.9 &-363.4&  1588.0 &-395.4&  -323.4&   -12.9\\
$$ SKM$^*$  &0.160&    -15.8&   216.6 &-386.1&  1768.9 &-430.6&  -349.0    &37.3\\
$$ SLy230b &0.160&    -16.0&   229.9 &-363.1&  1586.8 &-395.5&  -322.9&   -12.2\\
$$ SLy7  &0.158&  -15.9 &229.7    &-359.2&  1562.9 &-402.7&  -327.6&   -10.4\\
$$ SLy2  &0.160&  -15.9&   229.2 &-362.7&  1585.5  &-406.1&   -328.9&   -7.0\\
$$ SLy1  &0.160&   -16.0&   229.8 &-364.3&  1594.4 &-408.7&  -331.3&   -7.3\\
$$ SKM &0.160  &-15.8   &216.6    &-386.1&  1768.9 &-444.9&  -356.9&   45.1\\
$$ SII &0.148 &-16.0   &341.4    &15.8&   -567.5 &-565.9   &-568.2   &23.2\\
$$ SLy5  &0.161&  -16.0&   229.9 &-364.1&  1592.7 &-413.7&  -334.0&   -4.6\\
\hline\hline
\end{tabular}%
\end{table}
\begin{table}[tbp]
\caption{{\protect\small Continued with Table \ref{E0tab2}}}
\label{E0tab3}
\begin{tabular}{ccccccccc}
\hline\hline
Force & $\rho_0$ & $E_0(\rho_0)$ & \quad $K_0$ & \quad $J_0$ & \quad $I_0$ & $K_{asy}$ & $K_{sat,2}$ & $K_{sat,4}$ \quad \\
\hline
$$ SLy9  &0.151&    -15.8&  229.8&  -350.4& 1511.4& -413.7& -329.2& 1.7\\
$$ SkI6  &0.159&    -15.9&  248.2&  -326.6& 1251.8& -402.2& -324.3& 15.5\\
$$ SkI4  &0.160&    -15.9&  248.0&  -331.2& 1280.0& -402.9& -322.2& 21.9\\
$$ SIV &0.151   &-16.0& 324.6&  -68.8&  -234.9& -517.7& -504.2& 64.8\\
$$ SGI &0.154   &-15.9& 261.8&  -297.9& 1005.0& -435.2& -362.5& 61.8\\
$$ SKO$^*$&0.160&   -15.7&  222.1&  -390.2& 1706.7& -495.6& -373.2& 98.3\\
$$ SkMP  &0.157&    -15.6&  231.0&  -338.4& 1425.2& -468.9& -366.8& 101.5\\
$$ SKa &0.155   &-16.0& 263.2&  -300.1& 1014.4& -526.2& -441.1& 105.0\\
$$ SKO &0.160   &-15.8& 222.8&-391.3&   1712.3& -519.1& -379.5& 143.9\\
$$ R$_{\sigma}$&0.158&  -15.6&  237.4&  -348.4& 1377.2& -523.3& -397.5& 184.5\\
$$ G$_{\sigma}$ &0.158& -15.6& 237.2&   -348.7& 1379.5& -550.1& -411.9& 230.6\\
$$ SKT4  &0.159&    -16.0&  235.5&  -383.0& 1562.5& -589.3& -436.2& 201.2\\
$$ SV &0.155&   -16.0&  305.7&  -175.8& 183.5&  -552.4& -497.1& 176.2\\
$$ SkI3  &0.158&    -16.0&  258.2&  -303.9& 1088.3& -530.1& -411.8& 152.1\\
$$ SkI2  &0.158&    -15.8&  240.9&  -339.7& 1351.4& -555.3& -408.2& 226.9\\
$$ SkI5  &0.156&    -15.8&  255.8&  -302.0& 1083.7& -616.4& -463.7& 347.2\\

\hline\hline
\end{tabular}%
\end{table}

\begin{table}[tbp]
\caption{{\protect\small The characteristic parameters
$E_{\mathrm{sym}}(\rho _{0})$ (MeV), $L$ (MeV), $K_{\mathrm{sym}}$
(MeV), $J_{\mathrm{sym}}$ (MeV), $E_{\mathrm{sym,4}}(\rho _{0})$
(MeV), $L_{\mathrm{sym,4}}$ (MeV), $K_{\mathrm{sym,4}}$ (MeV),
$m_{s,0}^{\ast }/m$ and $m_{v,0}^{\ast }/m$ at saturation density
for the MDI interaction with $x=-1, 0, 1$ and the SHF predictions
with $63$ standard Skyrme interactions.}} \label{Esymtab1}
\begin{tabular}{cccccccccc}
\hline\hline
Force & $E_{sym}(\rho_0)$ & \quad $L$ & $K_{sym}$ & $J_{sym}$ & $E_{sym,4}(\rho_0)$ & $L_{sym,4}$ & $K_{sym,4}$ & $m_{s,0}^{*}/m$ & $m_{v,0}^{*}/m$ \quad \\
\hline
$$ MDI(1) & 30.5 & 14.7  &  -264.0 & 660.0 & 0.62 & 0.53 & -4.82 & 0.67 & 0.54 \\
$$ MDI(0) & 30.5 & 60.2  &  -81.7 & 295.3 & 0.62 & 0.53 & -4.82 & 0.67 & 0.54 \\
$$ MDI(-1) & 30.5 & 105.8  &  100.7 & -69.3 & 0.62 & 0.53 & -4.82 & 0.67 & 0.54 \\
$$Z&26.8&   -49.7&  -657.9& 495.2&  0.78&    2.56 &   1.47 & 0.84 & 0.73 \\
$$E$_{\sigma}$&26.4& -36.9&  -457.8& 880.0&  0.88&    3.04&    2.38&0.84&0.70\\
$$E&27.7&    -31.3&  -570.7& 448.6&  0.80&    2.66&    1.67&0.87&0.74\\
$$Z$_{\sigma}$&26.7&    -29.4&  -401.4& 883.1&  0.91&    3.14&    2.58&0.78&0.66\\
$$SVI & 26.9& -7.3&   -471.3& 146.0& 0.67& 2.08&    0.76&0.95&0.81\\
$$Z$_{\sigma}^*$&28.8&  -4.5&   -332.6& 725.1&  0.92&    3.24&    2.79&0.77&0.65\\
$$SkSC4&28.8&   -2.2&   -329.5& 708.3&  0.46&    0.91&    -1.85&1.00&1.00\\
$$SI&29.2&  1.2&    -461.8& 141.4&  0.70&    2.16&    0.73&0.91&0.80\\
$$BSk3&27.9&    6.8&    -306.9& 550.3&  0.71&    2.19&    0.76&1.12&0.89\\
$$BSk1&27.8&    7.2&    -281.8& 606.4&  0.43&    0.79&    -2.04&1.05&1.05\\
$$BSk2&28.0&    8.0&    -297.0& 557.9&  0.71&    2.18&    0.74&1.04&0.86\\
$$MSk7&27.9& 9.4&    -274.6& 592.1&  0.43&    0.79&    -2.04&1.05&1.05\\
$$SIII&28.1&    9.9&    -393.7& 130.4&  0.83&    2.89&    2.34&0.76&0.66\\
$$BSk4&28.0&    12.5&   -265.9& 558.4&  0.61&    1.70&    -0.22&0.92&0.85\\
$$BSk8&28.0&    14.9&   -220.9& 624.9&  0.43&    0.78&    -2.09&0.80&0.87\\
$$BSk6&28.0&    16.8&   -215.2& 603.5&  0.45&    0.89&    -1.85&0.80&0.86\\
$$BSk7&28.0&    18.0&   -209.4& 598.2&  0.42&    0.77&    -2.08&0.80&0.87\\
$$SKP&30.0& 19.6&   -266.8& 508.6&  0.94&    3.33&    2.96&1.00&0.74\\
$$BSk5&28.7&    21.4&   -240.3& 499.9&  0.64&    1.83&    0.04&0.92&0.84\\
$$SKXm&31.2&    32.1&   -242.8& 428.7&  0.88&    3.02&    2.40&0.97&0.75\\
$$RATP&29.2&    32.4&   -191.2& 440.6&  1.06&    3.94&    4.21&0.67&0.56\\
\hline\hline
\end{tabular}%
\end{table}
\begin{table}[tbp]
\caption{{\protect\small Continued with Table \ref{Esymtab1}}}
\label{Esymtab2}
\begin{tabular}{cccccccccc}
\hline\hline
Force & $E_{sym}(\rho_0)$ & \quad $L$ & $K_{sym}$ & $J_{sym}$ & $E_{sym,4}(\rho_0)$ & $L_{sym,4}$ & $K_{sym,4}$ & $m_{s,0}^{*}/m$ & $m_{v,0}^{*}/m$\quad \\
\hline
$$SKX&31.1& 33.2&   -252.1& 379.7&  0.89&    3.10&    2.61&0.99&0.75\\
$$ SKXce  &30.1&    33.5&   -238.4& 356.9&  0.89&    3.12&    2.65&1.01&0.75\\
$$ BSk15  &30.0&    33.6&   -194.4& 466.5&  0.64&    1.84 &0.03&0.80&0.77\\
$$ BSk16  &30.0&    34.9&   -187.4& 461.9&  0.60&    1.66&    -0.31&0.80&0.78\\
$$ BSk10  &30.0&    37.2&   -194.9& 397.0&  0.69&    2.08&    0.50&0.92&0.81\\
$$ SGII  &26.8& 37.6&   -145.9& 330.4&  0.87&    3.01&    2.39&0.79&0.67\\
$$ BSk12  &30.0&    38.0&   -191.4& 392.5&  0.68&    2.02&    0.40&0.92&0.82\\
$$ BSk11  &30.0&    38.4&   -189.8& 390.1&  0.67&    2.01&    0.37&0.92&0.82\\
$$ SLy10 &32.0& 38.7&   -142.2& 591.2&  0.37&    0.50&    -2.60&0.68&0.80\\
$$ BSk13  &30.0&    38.8&   -187.9& 386.6&  0.67&    2.00&    0.37&0.92&0.82\\
$$ BSk9  &30.0 &    39.9&   -145.3& 475.8&  0.37&    0.47&    -2.71&0.80&0.91\\
$$ BSk14  &30.0&    43.9&   -152.0& 388.3&  0.60&    1.65&    -0.33&0.80&0.78\\
$$ SLy230a &32.0&   44.3&   -98.2&  602.9&  0.06&    -1.07&   -5.80&0.70&1.00\\
$$ SLy6  &31.2& 45.2&   -112.5& 511.3&  0.38&    0.55&    -2.54&0.69&0.80\\
$$ SLy8  &31.4& 45.3&   -116.5& 511.4&  0.40&    0.64&    -2.39&0.70&0.80\\
$$ SLy4  &31.8& 45.4&   -119.9& 521.0&  0.40&    0.63&    -2.40&0.69&0.80\\
$$ SLy0  &31.5& 45.4&   -116.8& 510.6&  0.40&    0.64&    -2.39&0.70&0.80\\
$$ SLy3  &32.1& 45.5&   -122.1& 526.2&  0.40&    0.63&    -2.41&0.70&0.80\\
$$ SKM$^*$  &30.0&  45.8&   -155.9& 330.5&  0.94&    3.32&    2.97&0.79&0.65\\
$$ SLy230b &32.0&   46.0&   -119.7& 521.5&  0.40&    0.61&    -2.43&0.69&0.80\\
$$ SLy7  & 32.4&    48.1&   -114.3& 516.6&  0.38&    0.55&    -2.53&0.69&0.80\\
$$ SLy2  &32.3& 48.8&   -113.5& 502.9&  0.40&    0.63&    -2.39&0.70&0.80\\
$$ SLy1  & 32.5&    48.8&   -115.7& 508.5&  0.40&    0.63&    -2.40&0.70&0.80\\
$$ SKM &30.7    &49.3&  -148.8& 323.3&  0.91&    3.19&    2.71&0.79&0.66\\
$$ SII &34.2&   50.0&   -265.7& 104.7&  1.10&    4.21&    4.94&0.58&0.50\\
$$ SLy5  &32.7& 50.3&   -111.9&    499.2&  0.40&    0.63&    -2.40&0.70&0.80\\
\hline\hline
\end{tabular}%
\end{table}
\begin{table}[tbp]
\caption{{\protect\small Continued with Table \ref{Esymtab2}}}
\label{Esymtab3}
\begin{tabular}{cccccccccc}
\hline\hline
Force & $E_{sym}(\rho_0)$ & \quad $L$ & $K_{sym}$ & $J_{sym}$ & $E_{sym,4}(\rho_0)$ & $L_{sym,4}$ & $K_{sym,4}$& $m_{s,0}^{*}/m$ & $m_{v,0}^{*}/m$ \quad \\
\hline
$$ SLy9  &32.1& 55.4&   -81.3&  461.8&  0.33& 0.32&   -2.88&0.67&0.80\\
$$ SkI6  &29.9& 59.2&   -46.8&  378.1&  0.28&    0.04&    -3.55&0.64&0.80\\
$$ SkI4  &29.5& 60.4&   -40.6&  351.2&  0.30&    0.15&    -3.36&0.65&0.80\\
$$ SIV &31.2&   63.5&   -136.7& 79.5&   1.37&    5.51&    7.50&0.47&0.41\\
$$ SGI &28.3&   63.9&   -52.0&  194.5&  0.86&    3.00&    2.36&0.61&0.57\\
$$ SKO$^*$  &32.1&  69.7&   -77.5&  221.4&  0.55&    1.38&    -0.90&0.90&0.87\\
$$ SkMP  &29.7& 69.8&   -50.3&  159.7&  0.93&    3.31&    3.00&0.65&0.58\\
$$ SKa &32.9&   74.6&   -78.5&  174.5&  1.13&    4.33&    5.06&0.61&0.51\\
$$ SKO &32.0&   79.6&   -42.3&  130.0&  0.59&    1.57&    -0.53&0.89&0.85\\
$$ R$_{\sigma}$ &30.6&  85.7&   -9.1&   22.2&   0.85&    2.88&    2.13&0.78&0.68\\
$$ G$_{\sigma}$ &31.4&  94.0&   14.0&   -26.7&  0.85&    2.87&    2.12&0.78&0.68\\
$$ SKT4  &35.5& 94.1&   -24.5&  97.8&   0.45&    0.91&    -1.83&1.00&1.00\\
$$ SV &32.8&    96.1&   24.2&   48.0&   1.70&    7.18&    10.77&0.38&0.33\\
$$ SkI3  &34.8& 100.5&  73.0&   211.5&  0.12&    -0.74&   -5.10&0.58&0.82\\
$$ SkI2  &33.4& 104.3&  70.7&   51.6&   0.37&    0.48&    -2.66&0.68&0.80\\
$$ SkI5  &36.6& 129.3&  159.6&  11.7&   0.12&    -0.72&   -5.04&0.58&0.80\\
\hline\hline
\end{tabular}%
\end{table}
\end{widetext}

As shown in Section \ref{saturation}, the expressions for the
saturation density Eq. (\ref{rhosat}), the binding energy Eq.
(\ref{Esat}) and the isobaric incompressibility coefficient Eq.
(\ref{Ksat}) are exact up to $4$th-order in $\delta $, and these
expressions involve $11$ characteristic parameters defined at the
normal nuclear density $\rho _{0}$, i.e., $E_{0}(\rho _{0})$,
$K_{0}$, $J_{0} $, $I_{0}$, $E_{\mathrm{sym}}(\rho _{0})$, $L$,
$K_{\mathrm{sym}}$, $J_{\mathrm{sym}}$, $E_{\mathrm{sym,4}}(\rho
_{0})$, $L_{\mathrm{sym,4}}$, and $K_{\mathrm{sym,4}}$. We summarize
the values of $\rho _{0}$, $E_{0}(\rho _{0})$, $K_{0}$, $J_{0}$,
$I_{0}$, $K_{\mathrm{asy}}$, $K_{\mathrm{sat,2}}$ and
$K_{\mathrm{sat,2}}$ in Tables \ref{E0tab1}, \ref{E0tab2}, and
\ref{E0tab3} while the values of $E_{\mathrm{sym}}(\rho _{0})$, $L$,
$K_{\mathrm{sym}}$, $J_{\mathrm{sym}}$, $E_{\mathrm{sym,4}}(\rho
_{0})$, $L_{\mathrm{sym,4}}$, $K_{\mathrm{sym,4}}$, $m_{s,0}^{\ast
}/m$ and $m_{v,0}^{\ast }/m$ in Tabels \ref{Esymtab1},
\ref{Esymtab2}, and \ref{Esymtab3} for the MDI interaction with
$x=1,0,-1$ and the popular $63$ standard Skyrme interactions with
their saturation density and the symmetry energy satisfying $0.140$
fm$^{-3}<\rho _{0}<0.165$ fm$^{-3}$ and $25$
MeV$<E_{\mathrm{sym}}(\rho _{0})<37$ MeV, respectively. For the $63$
standard Skyrme interactions, the values in the Tables are sorted in
the order of increasing values of $L$. It should be stressed here
that the parameters of all Skyrme interactions are chosen to fit the
binding energies and charge radii of a large number of nuclei in the
periodic table. Detailed values of the parameters for these $63$
Skyrme interactions can be found in
Refs.~\cite{Bra85,Fri86,Bro98,Cha97,Cha98,Sto03,Sto07,Che05b,Sam02,Sam03,Gor03,Sam04,Gor05,Sam05,Gor06,Gor07,Cha08a,Gor08}.
The selected ranges of $\rho _{0}$ and $E_{\mathrm{sym}}(\rho _{0})$
are essentially consistent with their empirical values inferred from
experimental data. We note that here no constraints are imposed on
$K_{0}$ and $L$ for selecting the Skyrme interactions as we will
systematically explore the correlations of other physical quantities
with $K_{0}$ or $L$.

\begin{figure}[tbh]
\includegraphics[scale=1.2]{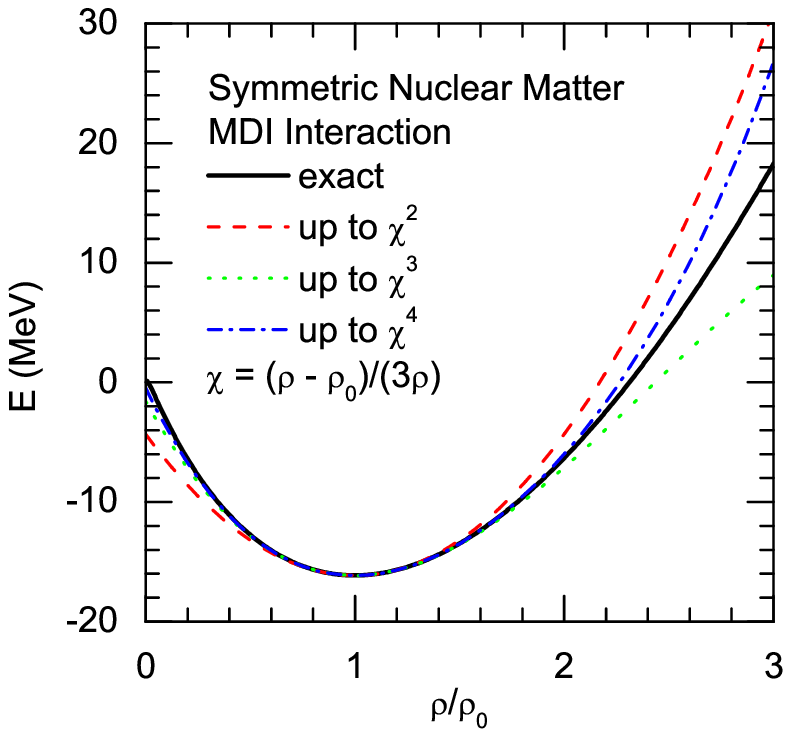}
\caption{{\protect\small (Color online) Energy per nucleon as a
function of density for symmetric nuclear matter in the MDI
interaction. Also included are results obtained by
using Eq. (\protect\ref{E0}) up to }$\protect\chi ^{2}${\protect\small , }$%
\protect\chi ^{3}${\protect\small , and }$\protect\chi ^{4}${\protect\small %
, respectively.}}
\label{EOS0MDIexpand}
\end{figure}

Since the $11$ characteristic parameters are defined at the normal
nuclear density $\rho _{0}$, it is of interest to study the extent
these characteristic parameters can provide information on the
properties of asymmetric nuclear at sub-saturation and
supra-saturation density regions. As an example, we show in Fig.
\ref{EOS0MDIexpand} the energy per nucleon of symmetry nuclear
matter from the MDI interaction as a function of its density. Also
shown in Fig. \ref{EOS0MDIexpand} are the results obtained by using
Eq. (\ref{E0}) including terms up to $\chi ^{2}$, $\chi ^{3}$, and
$\chi ^{4}$, respectively. It is seen that Eq. (\ref{E0}) with terms
up to $\chi ^{2}$, i.e., the parabolic approximation Eq.
(\ref{E0para}) which involves only two characteristic parameters,
i.e., $E_{0}(\rho _{0})$ and $K_{0}$, can approximate very well the
EOS of symmetric nuclear matter from about $0.5\rho _{0}$ to
$1.5\rho _{0}$. Including higher-order terms of $\chi ^{3}$ and
$\chi ^{4}$ with the characteristic parameters $J_{0}$ and $I_{0}$
in Eq. (\ref{E0}) improves significantly the approximation to the
EOS at low densities and that up to about $2\rho _{0}$. To describe
reasonably the EOS of symmetric nuclear matter above $2\rho _{0}$,
one needs to include higher-order terms in $\chi $ .

\begin{figure}[tbh]
\includegraphics[scale=1.2]{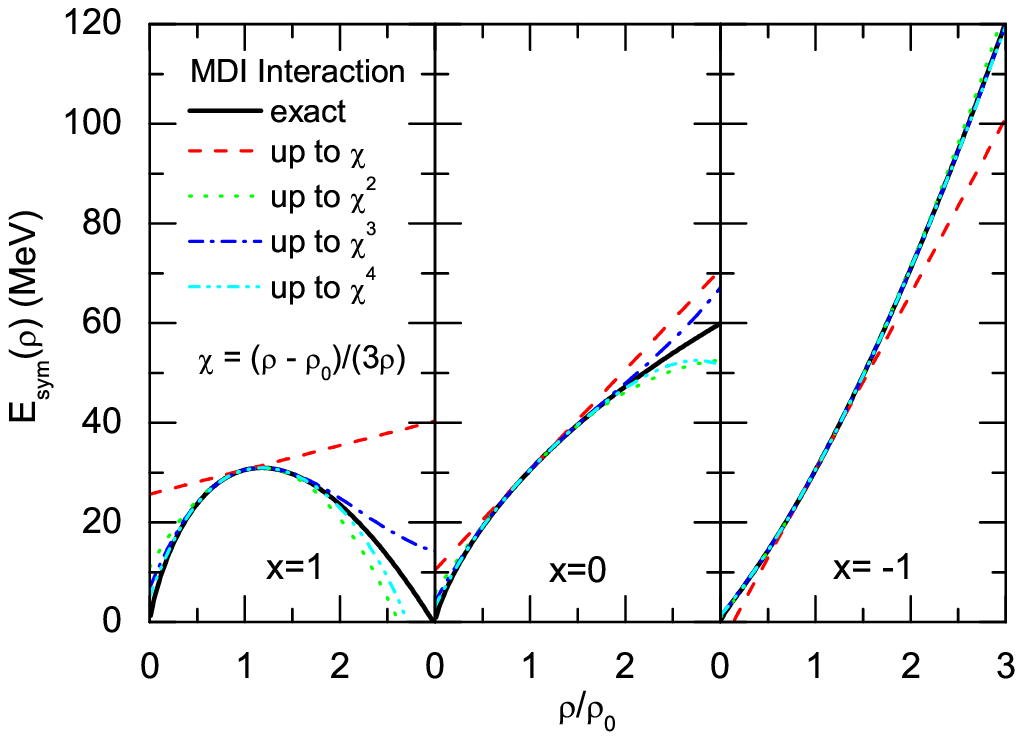}
\caption{{\protect\small (Color online) Density dependence of
symmetry energy using the MDI interaction with }$x=1${\protect\small
\ (left) }$0${\protect\small \ (middle), and }$-1${\protect\small \
(right) together with corresponding results obtained by using Eq.
(\protect\ref{EsymLKJI}) up
to }$\protect\chi ${\protect\small , }$\protect\chi ^{2}${\protect\small , }$%
\protect\chi ^{3}${\protect\small , and }$\protect\chi ^{4}${\protect\small %
, respectively.}}
\label{EsymMDIexpand}
\end{figure}

Fig. \ref{EsymMDIexpand} displays the density dependence of symmetry
energy using the MDI interaction with $x=1$, $0$, and $-1$ together
with corresponding results obtained by using Eq. (\ref{EsymLKJI}) up
to $\chi $, $\chi ^{2}$, $\chi ^{3}$, and $\chi ^{4}$, respectively.
It is seen that the importance of the contributions from
higher-order terms in $\chi $ to the density dependence of symmetry
energy depends on the stiffness of symmetry energy. For a supra-soft
symmetry energy ($x=1$), terms up to $\chi ^{3}$ are needed to
describe reasonably the symmetry energy from sub-saturation
densities to about $2\rho _{0}$. The situation is similar for the
case of modestly soft symmetry energy ($x=0$). For the stiffer
symmetry energy ($x=-1$), including terms up to $\chi ^{2}$ already
give a good description of the symmetry energy from sub-saturation
densities to about $3\rho _{0}$.

\begin{figure}[tbh]
\includegraphics[scale=1.2]{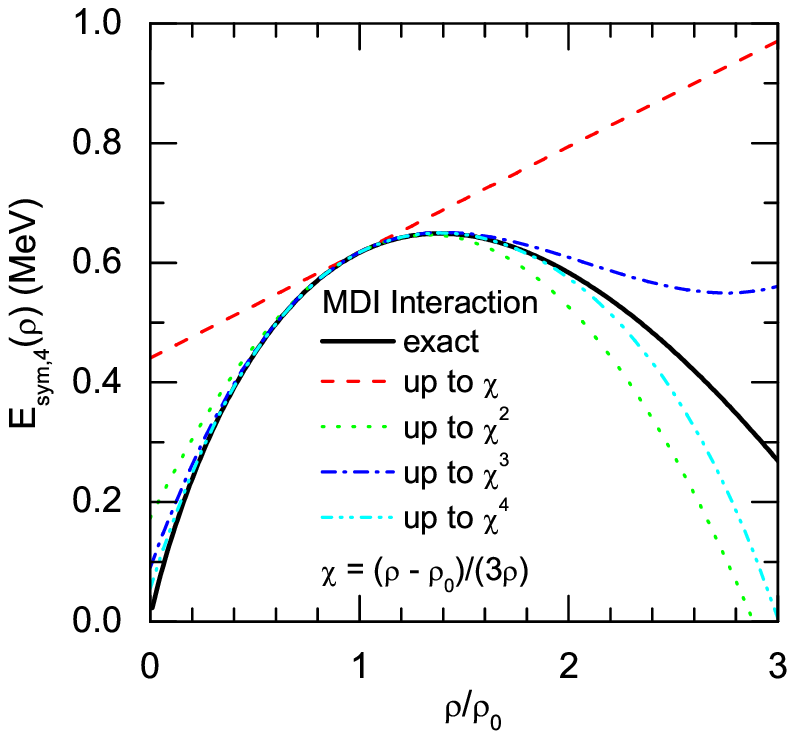}
\caption{{\protect\small (Color online) Density dependence of the }$4$%
{\protect\small th-order symmetry energy }$E_{\mathrm{sym,4}}(\protect\rho )$%
{\protect\small \ using the MDI interaction together with results
obtained by using Eq. (\protect\ref{Esym4LKJI}) up to }$\protect\chi $%
{\protect\small , }$\protect\chi ^{2}${\protect\small , }$\protect\chi ^{3}$%
{\protect\small , and }$\protect\chi ^{4}${\protect\small , respectively.}}
\label{Esym4MDIexpand}
\end{figure}

In Fig. \ref{Esym4MDIexpand}, we show the density dependence of the
$4$th-order symmetry energy $E_{\mathrm{sym,4}}(\rho )$ using the
MDI interaction together with corresponding results obtained by
using Eq. (\ref{Esym4LKJI}) up to $\chi $, $\chi ^{2}$, $\chi ^{3}$,
and $\chi ^{4}$, respectively. We note here that the
$E_{\mathrm{sym,4}}(\rho )$ does not depend on the $x$ parameter in
the MDI interaction as shown in Eq. (\ref{Esym4MDI}). Firstly, one
can see from Fig. \ref{Esym4MDIexpand} that the magnitude of
$E_{\mathrm{sym,4}}(\rho )$ is very small compared to that of
$E_{\mathrm{sym}}(\rho )$. As shown in Table \ref{Esymtab1}, the
value of $E_{\mathrm{sym,4}}(\rho )$ at the normal nuclear density
$\rho _{0}$ is about $0.62$ MeV which is consistent with the
predictions from the SHF approach using different Skyrme
interactions as shown in Tables \ref{Esymtab1}, \ref{Esymtab2}, and
\ref{Esymtab3} where one can see that only $5$ Skyrme interactions,
i.e., RATP, SII, SIV, SKa, and SV, among the $63$ Skyrme
interactions have $E_{\mathrm{sym,4}}(\rho _{0})$ larger than $1$
MeV (but still less than $2$ MeV). The value of
$E_{\mathrm{sym,4}}(\rho _{0})=0.62$ MeV is further consistent with
the value $0.57$ MeV predicted by the MSL model using Eq.
(\ref{Esym4MID}). These results thus confirm the empirical parabolic
law that the higher-order (including $4$th-order) contributions of
$\delta $ in the EOS of asymmetric nuclear matter are usually very
small and negligible as mentioned previously. Furthermore, similarly
to the cases shown in Figs. \ref{EOS0MDIexpand} and
\ref{EsymMDIexpand}, including terms up to $\chi ^{2} $ in Eq.
(\ref{Esym4LKJI}) can approximate very well the exact $4$-th-order
symmetry energy for the density region between about $0.5\rho _{0}$
and $1.5\rho _{0}$ while including higher-order terms of $\chi ^{3}$
and $\chi ^{4}$ improves significantly the results at low densities
and up to about $2\rho _{0}$.

The above results thus indicate that generally one needs
higher-order terms in $\chi $ (higher than $4$th-order) to describe
the EOS of asymmetric nuclear matter at high density region (above
$2\rho _{0}$). We note that above conclusions obtained with the MDI
interaction are also valid for the SHF approach and the MSL model.
These features imply that it is very difficult to obtain correct
information on the high density behaviors of the EOS for asymmetric
nuclear matter based on the characteristic parameters obtained at
the normal nuclear density $\rho _{0}$. At this point, it should be
mentioned that the transport model analysis of heavy-ion collisions
at intermediate and high energies as well as the astrophysical
observations, especially on the properties of compact stars, provide
unique tools to extract information on the EOS of asymmetric nuclear
matter at high densities \cite{Dan02a,Bar05,Fuc06a,Kla06,LCK08}.

\subsection{Isospin dependence of the saturation properties of
asymmetric nuclear matter}

In the following, we show the results on the saturation properties
of asymmetric nuclear matter, i.e., the saturation density as well
as the binding energy and incompressibility at saturation density.
Especially, we investigate their isospin dependence and study if the
higher-order terms in the isospin asymmetry $\delta $ ($\delta ^{4}$
term) are important for the description of the saturation properties
of asymmetric nuclear matter.

\subsubsection{The saturation density}

\begin{figure}[tbh]
\includegraphics[scale=1.2]{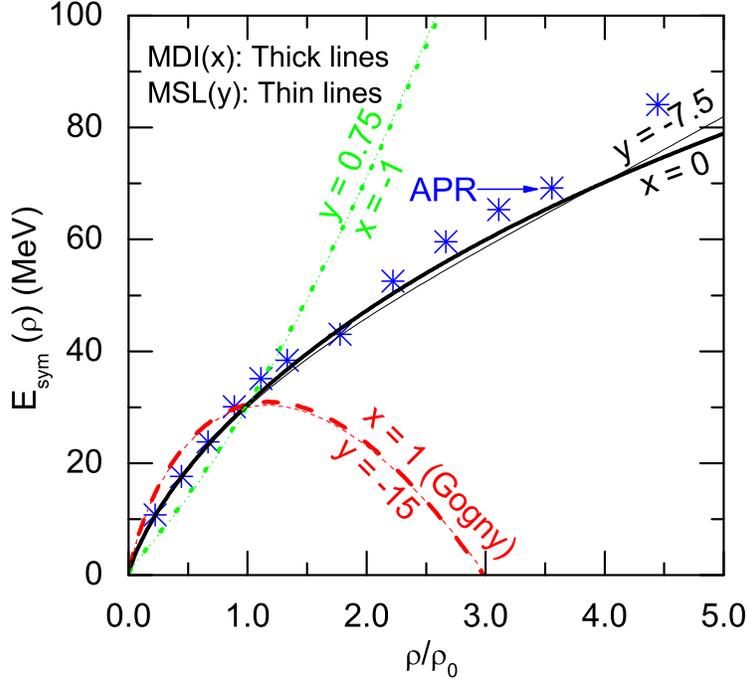}
\caption{{\protect\small (Color online) Density dependence of the symmetry
energy from the MDI interaction with }$x=1${\protect\small , }$0$%
{\protect\small , and }$-1${\protect\small . The results from the widely
used APR (Akmal-Pandharipande-Ravenhall) prediction \protect\cite{Akm98} and
the phenomenological MSL model prediction with }$y=-15${\protect\small , }$%
-7.5${\protect\small , and }$0.75${\protect\small \ are also included for
comparison.}}
\label{EsymRho}
\end{figure}
The saturation density is a basic quantity of asymmetric nuclear
matter. To see the symmetry energy dependence of the saturation
density of asymmetric nuclear matter, we use here the MDI
interaction with $x=1$, $0$, and $-1$. The density dependence of the
symmetry energy from this interaction is shown in Fig.
\ref{EsymRho}. Also included in Fig. \ref{EsymRho} are the results
from the widely used APR (Akmal-Pandharipande-Ravenhall) prediction
\cite{Akm98} and the phenomenological MSL model prediction with
$y=-15$, $-7.5$, and $0.75$ (We will discuss the MSL results later).
It is seen that the APR prediction for the symmetry energy resembles
very well that from the MDI interaction with $x=0$ up to about
$3.5\rho _{0}$.

\begin{figure}[tbh]
\includegraphics[scale=1.2]{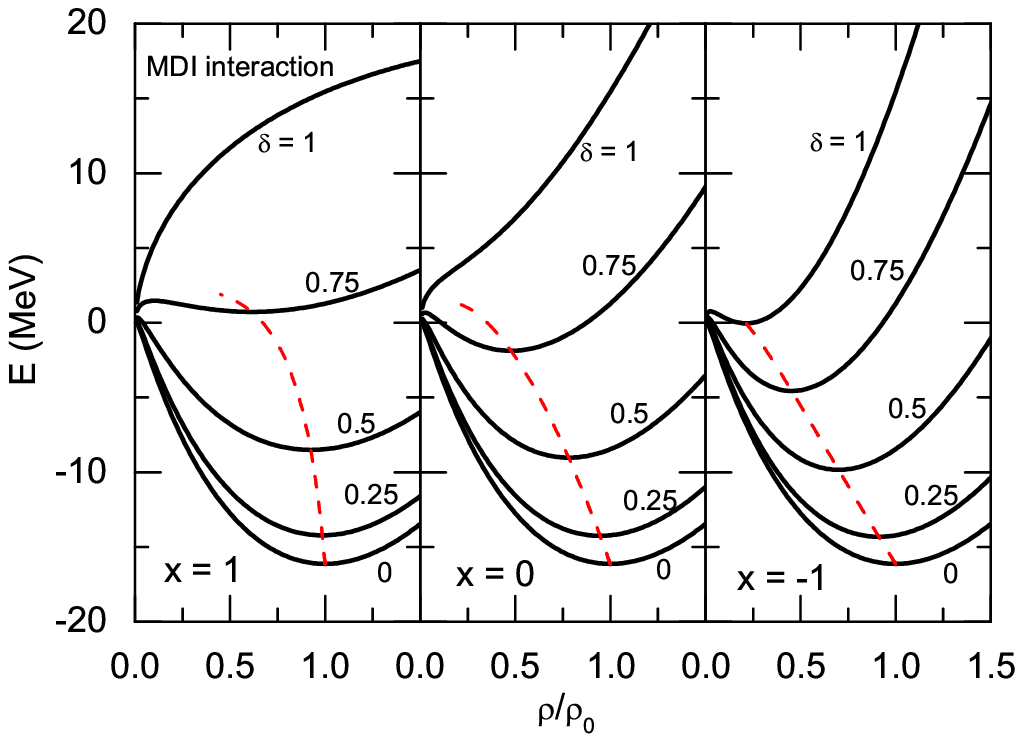}
\caption{{\protect\small (Color online) The density and isospin asymmetry
dependence of the binding energy per nucleon for asymmetric nuclear matter
in the MDI interaction with }$x=1${\protect\small , }$0${\protect\small ,
and }$-1${\protect\small . The saturation points at different isospin
asymmetries are also indicated.}}
\label{ErhoDelMDI}
\end{figure}

Using the MDI interaction with $x=1$, $0$, and $-1$, we have
calculated the density and isospin asymmetry dependence of the
binding energy per nucleon of asymmetric nuclear matter, and the
results are shown in Fig. \ref{ErhoDelMDI}. Further indicated in
Fig. \ref{ErhoDelMDI} are corresponding saturation points in the
$E$-$\rho $ plane. One can see that different symmetry energies lead
to different EOS of pure neutron matter ($\delta =1$) as expected.
In particular, the EOS of pure neutron matter for $x=-1 $ is bounded
at low densities. In addition, different symmetry energies lead to
rather different behaviors for the saturation points in the $E
$-$\rho $ plane.

\begin{figure}[tbh]
\includegraphics[scale=1.2]{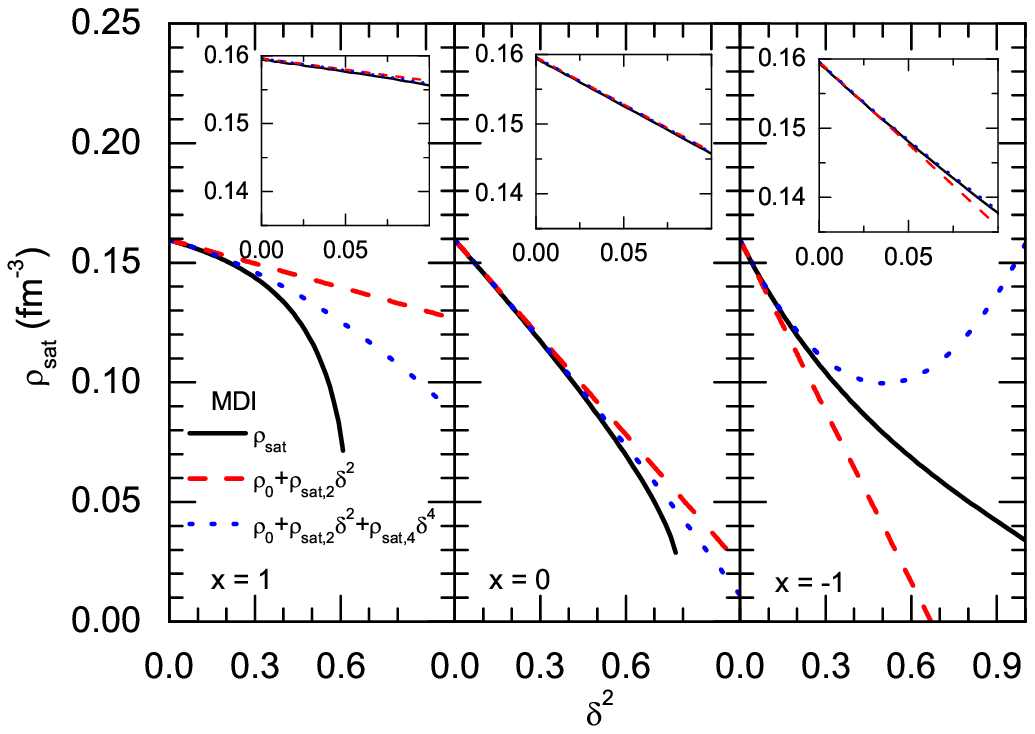}
\caption{{\protect\small (Color online) The saturation density }$\protect%
\rho _{\mathrm{sat}}${\protect\small \ as a function of
}$\protect\delta ^{2}${\protect\small \ in the MDI interaction with
}$x=1${\protect\small , }$0${\protect\small , and
}$-1${\protect\small . Corresponding results
from Eq. (\protect\ref{rhosat}) including terms up to }$\protect\delta ^{2}$%
{\protect\small \ and up to }$\protect\delta ^{4}${\protect\small ,
respectively, are also included for comparison. The inset in Fig.
\protect \ref{rhoSATDel2MDI} displays corresponding results at
smaller isospin asymmetries with }$\protect\delta ^{2}\leq
0.1${\protect\small .}} \label{rhoSATDel2MDI}
\end{figure}

In order to see more clearly the isospin dependence of the
saturation density, we show in Fig. \ref{rhoSATDel2MDI} the
saturation density $\rho _{\mathrm{sat}}(\delta )$ as a function of
$\delta ^{2}$ in the MDI interaction with $x=1$, $0$, and $-1$,
respectively. The exact saturation density $\rho
_{\mathrm{sat}}(\delta )$ is obtained from Eqs. (\ref{rhoSatDef})
and (\ref{EOSMDI}). Corresponding results from Eq. (\ref{rhosat})
including terms up to $\delta ^{2}$ and up to $\delta ^{4}$,
respectively, are also included for comparison. The results indicate
that the saturation density generally decreases with isospin
asymmetry and more neutron-rich nuclear matter has lower saturation
density. In addition, for the stiffer symmetry energy ($x=-1$), the
nuclear matter can be bounded even for pure neutron matter (The
corresponding saturation density is about $0.3\rho _{0}$). The inset
in Fig. \ref{rhoSATDel2MDI} displays corresponding results at
smaller isospin asymmetries with $\delta ^{2}\leq 0.1$ which is
relevant to the properties of finite nuclei. In the small isospin
asymmetry region ($\delta ^{2}\leq 0.1$), the saturation density
$\rho _{\mathrm{sat}}(\delta )$ displays a linear dependence on
$\delta ^{2} $ and Eq. (\ref{rhosat}) including terms up to $\delta
^{2}$ thus approximates very well the exact $\rho
_{\mathrm{sat}}(\delta )$. Furthermore, how $\rho
_{\mathrm{sat}}(\delta )$ decreases with increasing $\delta$ depends
on the stiffness of symmetry energy with the softer symmetry energy
having a weaker dependence, and this feature can be easily
understood from Eq. (\ref{rhosat2}) which indicates that the slope
of $\rho _{\mathrm{sat}}(\delta )$ with respect to $\delta ^{2}$ is
proportional to $-\frac{3L}{K_{0}}\rho _{0}$. In addition, at larger
isospin asymmetries with $\delta ^{2}\geq 0.3$, including
higher-order $\delta $ terms up to$\ \delta ^{4}$ in Eq.
(\ref{rhosat}) still deviates significantly from the exact $\rho
_{\mathrm{sat}}(\delta )$ and higher-order terms of $\delta $ are
thus necessary (except for the case of $x=0$ where Eq.
(\ref{rhosat}) with terms up to $\delta ^{4}$ gives a good
approximation to the exact $\rho _{\mathrm{sat}}(\delta )$ in the
whole $\delta $ region where the asymmetric matter can still be
bounded). These features imply that higher-order terms in $\delta $
may be important for the determination of the saturation density of
nuclear matter at very neutron-rich nuclear environment, such as
inside a neutron star.

\subsubsection{Binding energy at saturation density}

\begin{figure}[tbh]
\includegraphics[scale=1.2]{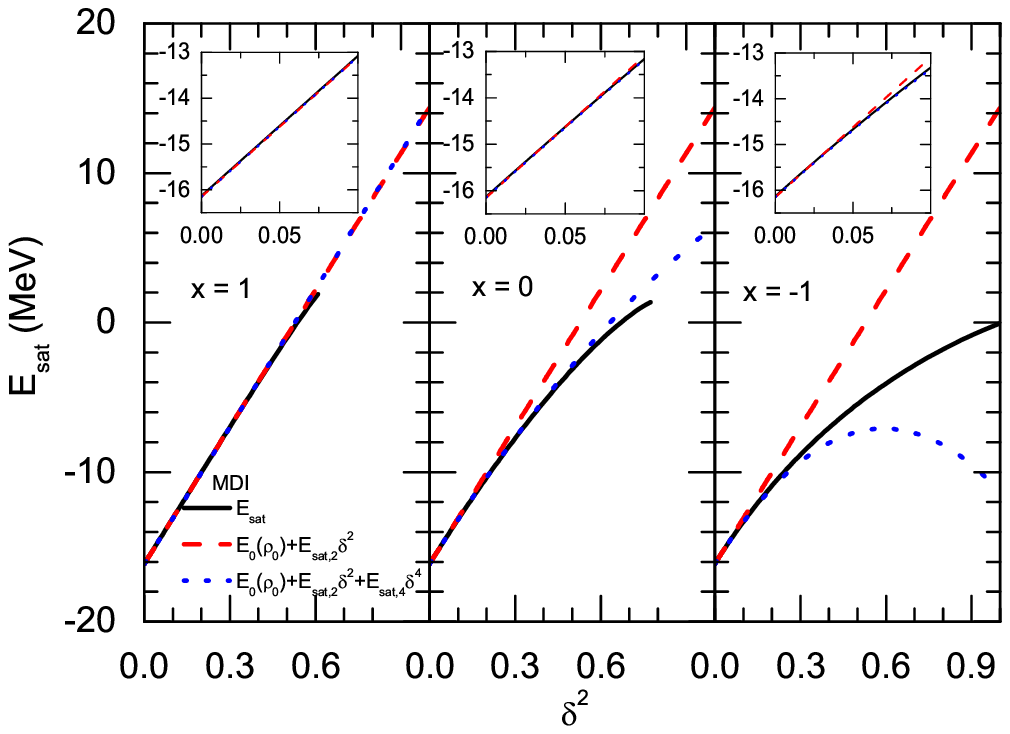}
\caption{{\protect\small (Color online) Same as Fig. \protect\ref%
{rhoSATDel2MDI} but for the binding energy at saturation density
$E_{\mathrm{sat}}$.}} \label{EsatDel2MDI}
\end{figure}

The isospin dependence of the binding energy per nucleon of
asymmetric nuclear matter at saturation density
$E_{\mathrm{sat}}(\delta )$ is shown in Fig. \ref{EsatDel2MDI} as a
function of $\delta ^{2}$ for the MDI interaction with $x=1$, $0$,
and $-1$. The exact $E_{\mathrm{sat}}(\delta )$ is obtained from
Eqs. (\ref{rhoSatDef}) and (\ref{EOSMDI}). Corresponding results
from Eq. (\ref{Esat}) including terms up to $\delta ^{2}$ and up to
$\delta ^{4}$, respectively, are also included for comparison. It is
seen that $E_{\mathrm{sat}}(\delta )$ generally increases with
increasing isospin asymmetry. The results at smaller isospin
asymmetries with $\delta ^{2}\leq 0.1$ are shown in the inset of
Fig. \ref{EsatDel2MDI} and it is seen that $E_{\mathrm{sat}}(\delta
)$ displays a linear dependence on $\delta ^{2}$ and therefore can
be very well approximated by Eq. (\ref{Esat}) including terms up to
$\delta ^{2}$. We note that the rate at which
$E_{\mathrm{sat}}(\delta )$ increases with $\delta ^{2}$ at small
$\delta ^{2}$ is determined uniquely by $E_{\mathrm{sym}}(\rho
_{0})$ as shown in Eq. (\ref{Esat}). Also, it is seen that including
higher-order $\delta $ terms up to $\ \delta ^{4}$ in Eq.
(\ref{Esat}) gives a good approximation to the exact
$E_{\mathrm{sat}}(\delta )$ in the whole $\delta $ region (except
for the case of $x=-1$ where Eq. (\ref{Esat}) including terms up to
$\delta ^{4}$ still deviates significantly from the exact
$E_{\mathrm{sat}}(\delta )$ at larger isospin asymmetries with
$\delta ^{2}\geq 0.3$ and higher-order terms of $\delta $ are thus
needed).

\subsubsection{Incompressibility at saturation density}

\begin{figure}[tbh]
\includegraphics[scale=1.2]{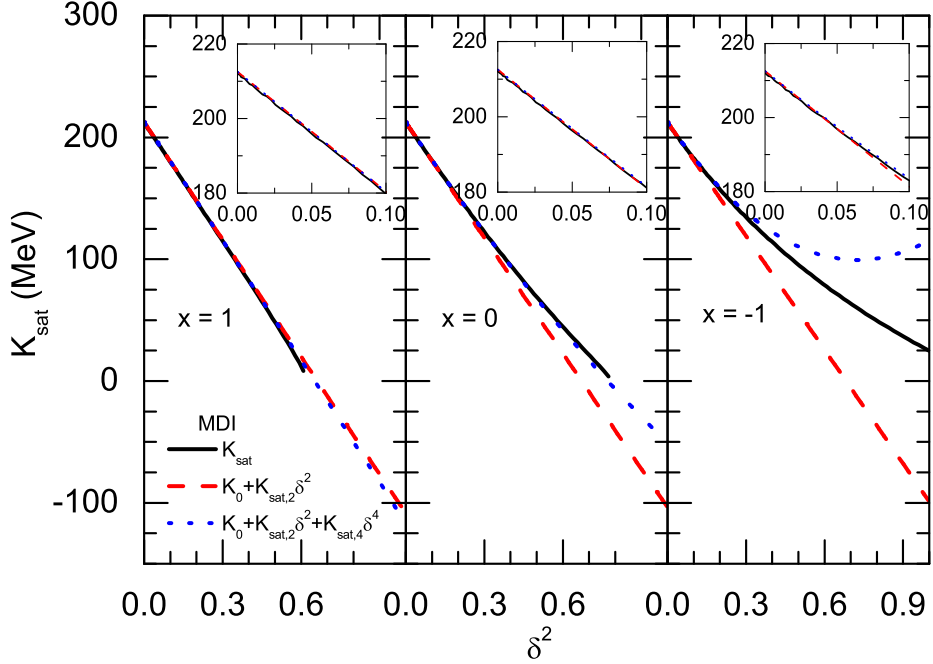}
\caption{{\protect\small (Color online) Same as Fig. \protect\ref%
{rhoSATDel2MDI} but for the incompressibility at saturation density
$K_{\mathrm{sat}}$.}} \label{KsatDel2MDI}
\end{figure}

Shown in Fig. \ref{KsatDel2MDI} is the incompressibility at
saturation density $K_{\mathrm{sat}}(\delta )$ as a function of
$\delta ^{2}$ for the MDI interaction with $x=1$, $0$, and $-1$. The
exact $K_{\mathrm{sat}}(\delta )$ is obtained from Eqs.
(\ref{rhoSatDef}), (\ref{KsatDef}), and (\ref{EOSMDI}), and
corresponding results from Eq. (\ref{Ksat}) including terms up to
$\delta ^{2}$ and up to $\delta ^{4}$ are also included for
comparison. It is seen that $K_{\mathrm{sat}}(\delta )$ generally
decreases with increasing isospin asymmetry and more neutron-rich
nuclear matter has smaller incompressibility. This feature is
consistent with earlier calculations based on microscopic many-body
approaches~\cite{Bom94}. The softening of the incompressibility of
asymmetric nuclear matter with increasing isospin asymmetry may have
important implications in understanding the mechanism for supernovae
explosions~\cite{Bom94,Bar85}. Corresponding results at smaller
isospin asymmetries with $\delta ^{2}\leq 0.1$ are given in the
inset of Fig. \ref{KsatDel2MDI}, and it shows that Eq. (\ref{Ksat})
including terms up to $\delta ^{2}$ approximates very well the exact
$K_{\mathrm{sat}}(\delta )$ as $K_{\mathrm{sat}}(\delta )$ displays
a good linear correlation with $\delta ^{2}$. As to the decreasing
rate of $K_{\mathrm{sat}}(\delta )$ with increasing $\delta ^{2}$ at
small $\delta ^{2}$, it is determined by the parameter
$K_{\mathrm{sat,2}}$ as shown in Eq. (\ref{Ksat}) which depends on
the characteristic parameters $J_{0}$, $K_{0}$, $L$, and
$K_{\mathrm{sym}}$. In addition, including higher-order $\delta $
terms up to$\ \delta ^{4}$ in Eq. (\ref{Ksat}) is seen to give a
good approximation to the exact $K_{\mathrm{sat}}(\delta )$ in the
whole $\delta $ region (except for the case of $x=-1$ where Eq.
(\ref{Ksat}) including terms up to $\delta ^{4}$ deviates
significantly from the exact $K_{\mathrm{sat}}(\delta )$ at larger
isospin asymmetries with $\delta ^{2}\geq 0.3$ and higher-order
terms of $\delta $ are thus important).

The above results indicate that the saturation properties of
asymmetric nuclear matter, i.e., the saturation density as well as
the binding energy and incompressibility at saturation density,
exhibit a good linear correlation with $\delta ^{2}$ at smaller
isospin asymmetries with $\delta ^{2}\leq 0.1$ which is relevant to
the properties of finite heavy nuclei. On the other hand, depending
on the stiffness of nuclear symmetry energy, higher-order terms in
$\delta $ ($\delta ^{4}$ and higher-order terms) may become
important for describing reasonably the saturation properties of
asymmetric nuclear matter at larger isospin asymmetries with $\delta
^{2}\geq 0.3$. The importance of higher-order isospin asymmetry
terms for the stiffer symmetry energy has also been observed in
previous studies on the transition density in neutron
stars~\cite{Xu09a,Xu09b}. In addition, the saturation density and
the incompressibility at saturation density generally decrease with
the magnitude of isospin asymmetry while the binding energy at
saturation density shows an opposite behavior. Again, we note that
above conclusions obtained from the MDI interaction are also valid
for the SHF approach and the MSL model. Our results are further
consistent with the very recent study based on the RMF\ model
\cite{Pie09}.

\subsection{Constraining the $K_{\mathrm{sat,2}}$ parameter from the
phenomenological MSL model}

\subsubsection{General information on the $K_{\mathrm{sat,2}}$ parameter}

\begin{figure}[tbh]
\includegraphics[scale=1.2]{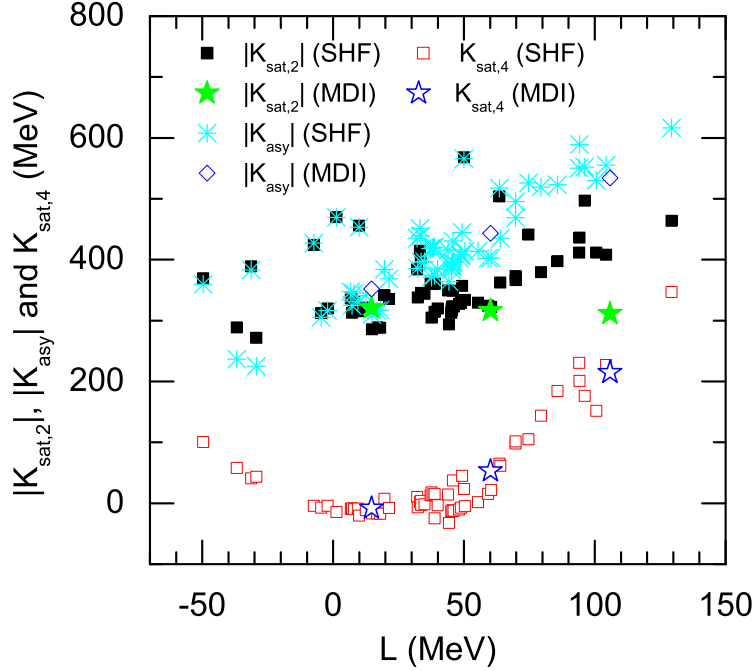}
\caption{{\protect\small (Color online) The absolute values of
}$K_{\mathrm{sat,2}}${\protect\small \ and
}$K_{\mathrm{asy}}${\protect\small \ and the value of
}$K_{\mathrm{sat,4}}${\protect\small \ as functions of
}$L${\protect\small \ for the MDI interaction with
}$x=1,0,-1${\protect\small \ and the }$63${\protect\small \ standard
Skyrme interactions considered in the present work.}}
\label{Ksat24KasyL}
\end{figure}
As shown in the above, the $K_{\mathrm{sat,2}}$ parameter
essentially characterizes the isospin dependence of the
incompressibility of asymmetric nuclear matter at saturation density
and the higher-order parameters (e.g., $K_{\mathrm{sat,4}}$) are
only important for extremely neutron-rich (or proton-rich) nuclear
matter with stiffer nuclear symmetry energies. Actually, it can be
seen from Tables \ref{E0tab1}, \ref{E0tab2}, and \ref{E0tab3} that
the magnitude (absolute values) of $K_{\mathrm{sat,2}}$ is generally
much larger than that of $K_{\mathrm{sat,4}}$ for the MDI
interaction with $x=1,0,-1$ and the $63$ standard Skyrme
interactions considered in the present work. Shown in Fig.
\ref{Ksat24KasyL} are the absolute values of $K_{\mathrm{sat,2}}$
and $K_{\mathrm{asy}}$ as well as the value of $K_{\mathrm{sat,4}}$
as functions of $L$ for the MDI interaction with $x=1$, $0$, $-1$
and the $63$ standard Skyrme interactions considered in the present
work. It is seen that these values of $K_{\mathrm{sat,2}}$ can be
nicely expressed as $-400\pm 120$ MeV. For the magnitude of the
$K_{\mathrm{asy}}$ parameter, it is generally larger than that of
the $K_{\mathrm{sat,2}}$ parameter, especially for the stiffer
symmetry energies (larger $L$ values), which indicates that the
higher-order $J_{0}$ parameter is important as discussed later.
Furthermore, the absolute values of $K_{\mathrm{sat,2}}$ are clearly
much larger than that of $K_{\mathrm{sat,4}}$ except that at very
large $L$ values the absolute values of $K_{\mathrm{sat,4}}$ may
become larger and comparable with that of $K_{\mathrm{sat,2}}$. This
feature is consistent with the results shown in
Fig.~\ref{KsatDel2MDI} where the higher-order terms are seen to be
only important for the stiffer symmetry energies.

It is generally believed that information on $K_{\mathrm{sat,2}}$
can be extracted experimentally by measuring the GMR in neutron-rich
nuclei~\cite{Bla80}. Usually, one can define a finite nucleus
incompressibility $K_{A}(N,Z)$ for a nucleus with $N$ neutrons and
$Z$ protons ($A=N+Z$) by the energy of GMR $E_{\mathrm{GMR}}$, i.e.,
\begin{equation}
E_{\mathrm{GMR}}=\sqrt{\frac{\hbar ^{2}K_{A}(N,Z)}{m\left\langle
r^{2}\right\rangle }},
\end{equation}%
where $m$ is the nucleon mass and $\left\langle r^{2}\right\rangle $
is the mean square mass radius of the nucleus in the ground state.
Similarly to the semi-empirical mass formula, the finite nucleus
incompressibility $K_{A}(N,Z) $ can be expanded as \cite{Bla80}
\begin{eqnarray}
K_{A}(N,Z) &=&K_{0}+K_{\mathrm{surf}}A^{-1/3}+K_{\mathrm{curv}}A^{-2/3}
\notag \\
&&+(K_{\tau }+K_{\mathrm{ss}}A^{-1/3})\left( \frac{N-Z}{A}\right) ^{2}
\notag \\
&&+K_{\mathrm{Coul}}\frac{Z^{2}}{A^{4/3}}+\cdot \cdot \cdot ,  \label{KA1}
\end{eqnarray}%
Neglecting the $K_{\mathrm{curv}}$ term, the $K_{\mathrm{ss}}$ term and
other higher-order terms in Eq. (\ref{KA1}), one can express the finite
nucleus incompressibility $K_{A}(N,Z)$ as
\begin{eqnarray}
K_{A}(N,Z)&=&K_{0}+K_{\mathrm{surf}}A^{-1/3}+K_{\tau }\left( \frac{N-Z}{A}%
\right) ^{2}\notag\\
&&+K_{\mathrm{Coul}}\frac{Z^{2}}{A^{4/3}}, \label{KA2}
\end{eqnarray}%
where $K_{0}$, $K_{\mathrm{surf}}$, $K_{\tau }$, and
$K_{\mathrm{coul}}$ represent the volume, surface, symmetry, and
Coulomb terms, respectively. The $K_{\tau }$ parameter is usually
thought to be equivalent to the $K_{\mathrm{sat,2}}$ parameter.
However, we would like to stress here that the $K_{\mathrm{sat,2}}$
parameter is a theoretically well-defined physical property of
asymmetric nuclear matter as shown previously while the value of the
$K_{\tau }$ parameter may depend on the details of the truncation
scheme in Eq. (\ref{KA1}). As shown in Ref.~\cite{Bla81}, $K_{\tau
}$ may be related to the isospin-dependent part of the surface
properties of finite nuclei, especially the surface symmetry energy.
Therefore, cautions are needed to interpret the $K_{\tau }$
parameter as the $K_{\mathrm{sat,2}}$ parameter (we will go back to
this point later).

Earlier attempts based on the above method have given widely
different values for the $K_{\tau }$ parameter. For example, a value
of $K_{\tau }=-320\pm 180$ MeV with a large uncertainty was obtained
in Ref. \cite{Sha88} from a systematic study of the GMR in the
isotopic chains of Sn and Sm. In this analysis, the value of $K_{0}$
was found to be $300\pm 25$ MeV, which is somewhat larger than the
commonly accepted value of $240\pm 20$ MeV. In a later study, an
even less stringent constraint of $-566\pm 1350<K_{\tau }<139\pm
1617$ MeV was extracted from the GMR of finite nuclei, depending on
the mass region of nuclei and the number of parameters used in
parameterizing the incompressibility of finite nuclei \cite{Shl93}.
More recently, a much more stringent constraint of $K_{\tau
}=-550\pm 100$ MeV has been obtained in Ref. \cite{LiT07,Gar07} from
measurements of the isotopic dependence of the GMR in even-A Sn
isotopes.

\subsubsection{Correlation between $J_{0}$ and $K_{0}$}

As shown in Eq. (\ref{Ksat2}), the $K_{\mathrm{sat,2}}$ parameter is
completely determined by the $4$ characteristic parameters $K_{0}$,
$J_{0}$, $L$, and $K_{\mathrm{sym}}$ at the normal nuclear density.
It thus would be interesting to estimate the possible value of
$K_{\mathrm{sat,2}}$ from knowledge on $K_{0}$, $J_{0}$, $L$, and
$K_{\mathrm{sym}}$. For the incompressibility of symmetric nuclear
matter at its saturation density $\rho _{0}$, the transport model
analyses on experimental data from subthreshold kaon production in
heavy-ion collisions favor a soft equation of state
\cite{Fuc01,Fuc06a,Har06}. More recently, the value of $K_{0}$ has
been more stringently determined to be $240\pm 20$ MeV from the
nuclear GMR data \cite{You99,Shl06,LiT07,Gar07,Col09}.

\begin{figure}[tbh]
\includegraphics[scale=1.5]{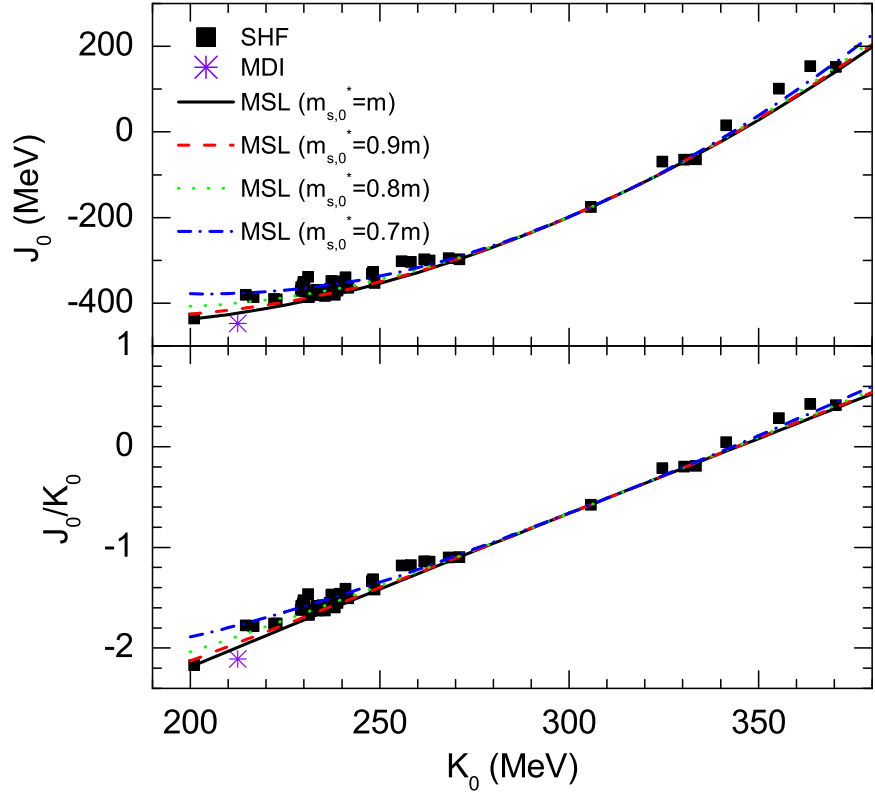}
\caption{{\protect\small (Color online) } The third derivative
parameter $J_{0}${\protect\small \ and } its ratio
$J_{0}/K_{0}${\protect\small \ to the incompressibility $K_0$ of
symmetry nuclear matter at saturation density} as functions of
$K_{0}${\protect\small \ from the MSL model, the MDI interaction and
the SHF prediction with the }$63${\protect\small \ Skyrme
interactions shown in Tables \protect\ref{E0tab1},
\protect\ref{E0tab2}, and \protect\ref{E0tab3}.}}
\label{J0K0abcSHFMDI}
\end{figure}

While the $K_{0}$ parameter has been relatively well determined, the
$J_{0}$ parameter is poorly known, and there is actually no
experimental information on the $J_{0}$ parameter. In the MSL model,
one can easily calculate from Eq. (\ref{E0MID}) the $J_{0}$
parameter as

\begin{eqnarray}
J_{0} &=&27\rho _{0}^{3}\frac{\partial ^{3}E_{0}(\rho )}{\partial ^{3}\rho }
|_{\rho =\rho _{0}}  \notag \\
&=&\frac{1}{E_{\rm kin}^{0}-3E_{0}-2C}[K_{0}^{2}+(18E_{0}-2E_{\rm
kin}^{0}-8C)K_{0} \notag \\
&&+12E_{\rm kin}^{0}E_{0}+6C(3E_{\rm kin}^{0}-25E_{0})].  \label{J0}
\end{eqnarray}%
In particular, for $m_{s,0}^{\ast }=0.8m$, we have
\begin{equation}
J_{0}=\frac{1}{59.1}\left( K_{0}^{2}-376.4K_{0}+11214.645\right) \text{
(MeV),}  \label{J0ms08}
\end{equation}%
while we have
\begin{equation}
J_{0}=\frac{1}{70.1}\left( K_{0}^{2}-332.2K_{0}-4243.2\right) \text{ (MeV),}
\label{J0ms1}
\end{equation}
for $m_{s,0}^{\ast }=m$ (and then $C=0$) and%
\begin{equation}
J_{0}=\frac{1}{51.2}\left( K_{0}^{2}-408K_{0}+22252\right) \text{ (MeV)}
\label{J0ms07}
\end{equation}%
for $m_{s,0}^{\ast }=0.7m$ and $C=9.47$ MeV. In above equations, the
unit of $K_{0}$ is MeV. Therefore, the $J_{0}$ parameter in the
phenomenological MSL model is a quadratic function of the $K_{0}$
parameter.

Shown in Fig. \ref{J0K0abcSHFMDI} are $J_{0}$ and $J_{0}/K_{0}$ as
functions of $K_{0}$ in the MSL model for different values of
$m_{s,0}^{\ast }=m$, $0.9m$, $0.8m$ and $0.7m$. Also included in
Fig. \ref{J0K0abcSHFMDI} are corresponding results from the MDI
interaction and the SHF prediction with the $63$ Skyrme interactions
shown in Tables \ref{E0tab1}, \ref{E0tab2}, and \ref{E0tab3}. It is
seen that the correlation between $J_{0}$ and $K_{0}$ is similar
among these three different models. Also, all three models show an
approximately linear correlation between $J_{0}/K_{0}$ and $K_{0}$.
This linear correlation can be easily understood from Eqs.
(\ref{J0ms08}), (\ref{J0ms1}) and (\ref{J0ms07}). For example, on
the r.h.s of Eq. (\ref{J0ms08}), the last term is small compared
with the first two terms, and thus one has $J_{0}\approx
\frac{1}{59.1}\left( K_{0}^{2}-376.4K_{0}\right) $, and then
$J_{0}/K_{0}\approx \frac{1}{59.1}\left( K_{0}-376.4\right) $ with
the units of $K_{0}$ in MeV. The linear dependence of $J_{0}/K_{0}$
on $K_{0}$ becomes better for larger values of $m_{s,0}^{\ast }$
which usually leads to smaller values for the last two terms on the
r.h.s of Eq. (\ref{J0}) as shown in Eq. (\ref{J0ms1}). We note that
the correlation between $J_{0}$ and $K_{0}$ obtained in the present
work is also consistent with the early finding by Pearson
\cite{Pea91}. In addition, it is seen from Fig. \ref{J0K0abcSHFMDI}
that the $m_{s,0}^{\ast }$ only has visible effects on the
correlation between $J_{0}$ and $J_{0}/K_{0}$ for smaller $K_{0}$
values while it has almost no effects for larger $K_{0}$ values.
While there do not exist any empirical constraints on the $J_{0}$
parameter, we assume in the present study the correlation between
$J_{0}$ and $K_{0}$ from the MSL model is valid and then determine
$J_{0}/K_{0}$ from the experimental constraint on $K_{0}$.

\subsubsection{Correlation between $L$ and $K_{\mathrm{sym}}$}

\begin{figure}[tbh]
\includegraphics[scale=1.2]{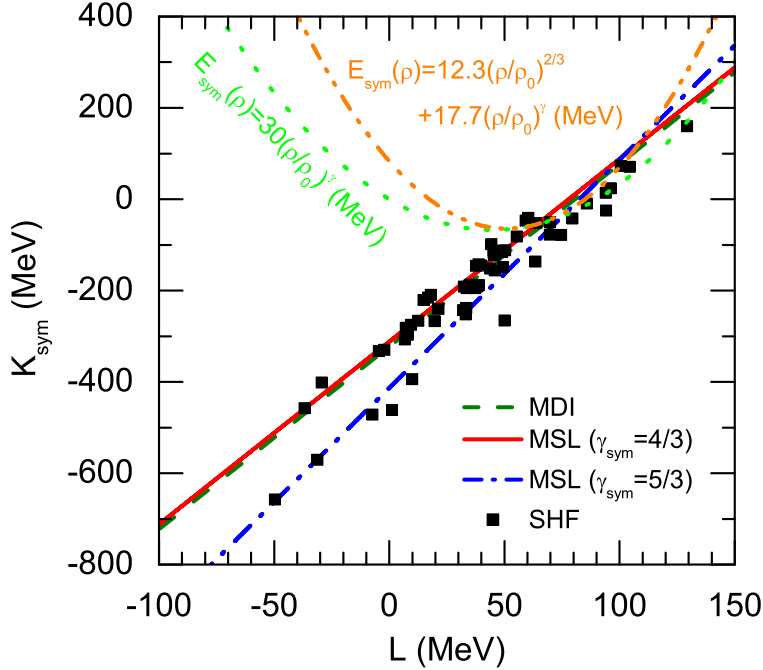}
\caption{{\protect\small (Color online) Correlation between\ }$K_{\mathrm{sym%
}}${\protect\small \ and }$L${\protect\small \ from the MSL model with }$%
\protect\gamma _{\mathrm{sym}}=4/3${\protect\small \ and }$5/3$%
{\protect\small , the MDI interaction and the SHF prediction with the }$63$%
{\protect\small \ Skyrme interactions shown in Tables
\protect\ref{Esymtab1}, \protect\ref{Esymtab2}, and
\protect\ref{Esymtab3}. The results from two simple one-parameter
symmetry energies in Eqs. (\protect\ref{EsymSimp1}) and
(\protect\ref{EsymSimp2}) are also shown for comparison.}}
\label{KsymL}
\end{figure}
The parameters $L$ and $K_{\mathrm{sym}}$ are determined by the
density dependence of the symmetry energy around saturation density.
In recent years, significant progress has been made both
experimentally and theoretically in extracting the information on
the behaviors of nuclear symmetry energy at sub-saturation density.
Using the isospin- and momentum-dependent IBUU04 transport model
with in-medium NN cross sections, the isospin diffusion data were
found to be consistent with the symmetry energy from the MDI
interaction with $x$ between $0$ and $-1$, which can be
parameterized by a density-dependent symmetry energy of $E_{\mathrm{sym}%
}(\rho )\approx 31.6(\rho /\rho _{0})^{\gamma }$ with $\gamma =0.69-1.05$ at
subnormal density ($\rho \leq \rho _{0}$) \cite{Che05a,LiBA05c,Che05b},
leading to the extraction of $61$ MeV $\leq L\leq 111$ MeV and $-82$ MeV $%
\leq K_{\mathrm{sym}}\leq 101$ MeV \cite{Che05a,LiBA05c,Che05b}. Using the
Skyrme interactions consistent with the EOS obtained from the MDI
interaction with $x$ between $0$ and $-1$, the neutron-skin thickness of
heavy nuclei calculated within the Hartree-Fock approach is consistent with
available experimental data \cite{Che05b,Ste05b} and also that from a
relativistic mean-field model based on an accurately calibrated parameter
set that reproduces the GMR in $^{90}$Zr and $^{208}$Pb as well as the
isovector giant dipole resonance of $^{208}$Pb \cite{Tod05}. The extracted
symmetry energy further agrees with the symmetry energy $E_{\mathrm{sym}%
}(\rho )=31.6(\rho /\rho _{0})^{0.69}$ recently obtained from the isoscaling
analyses of isotope ratios in intermediate energy heavy ion collisions \cite%
{She07}, which gives $L\approx 65$ MeV and $K_{\mathrm{sym}}\approx -61$
MeV. Furthermore, the above limited range of $E_{\mathrm{sym}}(\rho )$ at
subsaturation density is essentially consistent with the symmetry energy $E_{%
\mathrm{sym}}(\rho )=12.5(\rho /\rho _{0})^{2/3}+17.6(\rho /\rho
_{0})^{\gamma }$ with $\gamma =0.4-1.05$, extracted very recently
from the ImQMD (Improved QMD) model analyses of both the isospin
diffusion data and the double neutron/proton ratio \cite{Tsa09}. The
symmetry energy $E_{\mathrm{sym}}(\rho )=12.5(\rho /\rho
_{0})^{2/3}+17.6(\rho /\rho _{0})^{\gamma }$ with $\gamma =0.4-1.05$
leads to the constraints of $46$ MeV $\leq L\leq 80$ MeV and $-63$
MeV $\leq K_{\mathrm{sym}}\leq -17$ MeV.

It should be noted that all above constraints on $L$ and
$K_{\mathrm{sym}}$ are based on some unique energy density
functionals and thus special correlation between $L$ and
$K_{\mathrm{sym}}$ has been implicitly assumed. It is thus of
interest to study if there exists a universal correlation between
$L$ and $K_{\mathrm{sym}}$ in different models. For the MDI
interaction, it is implied from Eq. (\ref{EsymMDI}) that the $L$ and
$K_{\mathrm{sym}}$ both change linearly with the parameter $x$.
Therefore they are linearly correlated by varying the parameter $x$
that changes the density dependence of the symmetry energy. For the
MSL model, we can obtain from Eq. (\ref{EsymMID}) following
expressions
\begin{eqnarray}
L &=&2E_{\text{\textrm{sym}}}^{kin}({\rho _{0}})+5D+3\left[ E_{\text{\textrm{%
sym}}}({\rho _{0}})-E_{\text{\textrm{sym}}}^{kin}({\rho _{0}})-D\right]
\notag \\
&&+3y(\gamma _{\mathrm{sym}}-1)\left[ E_{\text{\textrm{sym}}}({\rho _{0}}%
)-E_{\text{\textrm{sym}}}^{kin}({\rho _{0}})-D\right]  \label{LMID} \\
K_{\text{\textrm{sym}}} &=&9y\gamma _{\mathrm{sym}}(\gamma _{\mathrm{sym}}-1)%
\left[ E_{\text{\textrm{sym}}}({\rho _{0}})-E_{\text{\textrm{sym}}}^{kin}({%
\rho _{0}})-D\right]  \notag \\
&&+10D-2E_{\text{\textrm{sym}}}^{kin}({\rho _{0}}).  \label{KsymMID}
\end{eqnarray}%
Therefore, for the MSL model, Eq. (\ref{LMID}) and Eq.
(\ref{KsymMID}) show that the $L$ and $K_{\mathrm{sym}}$ both change
linearly with the parameter $y$, and thus they are also linearly
correlated by varying the parameter $y$ to change the density
dependence of the symmetry energy. In particular, we have
\begin{eqnarray}
K_{\text{\textrm{sym}}} &=&3\gamma _{\mathrm{sym}}L+E_{\text{\textrm{sym}}%
}^{kin}({\rho _{0}})(3\gamma _{\mathrm{sym}}-2)  \notag \\
&&+2D(5-3\gamma _{\mathrm{sym}})-9\gamma _{\mathrm{sym}}E_{\text{\textrm{sym}%
}}({\rho _{0}}).  \label{KsymLMID}
\end{eqnarray}%
Also, $L$ and $K_{\mathrm{sym}}$ are expected to be correlated
within the SHF energy density functional.

Shown in Fig. \ref{KsymL} are the correlation between
$K_{\mathrm{sym}}$ and $L$ from the MSL model with $\gamma
_{\mathrm{sym}}=4/3$ and $5/3$ (We used here the default values of
$E_{\text{\textrm{sym}}}^{kin}({\rho _{0}})=12.3$ MeV,
$E_{\text{\textrm{sym}}}({\rho _{0}})=30$ MeV, and $D=-3.51$ MeV
from $m_{s,0}^{\ast }=0.8m$ and $m_{v,0}^{\ast }=0.7m$), the MDI
interaction and the SHF prediction with the $63$ Skyrme interactions
shown in Tables \ref{Esymtab1}, \ref{Esymtab2}, and \ref{Esymtab3}.
It is seen that the $K_{\mathrm{sym}}$ parameter indeed displays
approximately a linear correlation with the $L$ parameter for the
SHF predictions with the $63 $ Skyrme interactions and this linear
correlation is nicely reproduced by the MDI interaction and the MSL
model with $\gamma _{\mathrm{sym}}=4/3$. For the MSL model, one can
see from Eq. (\ref{KsymLMID}) that the $\gamma _{\mathrm{sym}} $
parameter controls the shape (slope) of the linear correlation
between $L $ and $K_{\mathrm{sym}}$. Furthermore, Fig. \ref{KsymL}
shows that results from some Skyrme interactions deviate from the
linear correlation obtained by the MDI interaction and the MSL model
with $\gamma _{\mathrm{sym}}=4/3$. In order to consider the
uncertainty in the shape (slope) for the correlation between $L$ and
$K_{\mathrm{sym}}$, we also include the result from the MSL model
with $\gamma _{\mathrm{sym}}=5/3$. The correlation between
$K_{\mathrm{sym}}$ and $L$ from the SHF predictions with the $63$
Skyrme interactions is consistent nicely with that from the MSL
model with $\gamma _{\mathrm{sym}}=4/3$ and $5/3$. Furthermore, we
find that the isospin-dependent nucleon effective mass has very
small effects on the correlation between $K_{\mathrm{sym}}$ and $L$.
This can be easily understood from Eq. (\ref{KsymLMID}) since the
value of $2D(5-3\gamma _{\mathrm{sym}})$ is only about $7$ MeV for
$\gamma _{\mathrm{sym}}=4/3$ (it is zero for $\gamma
_{\mathrm{sym}}=5/3$). The linear correlation between
$K_{\mathrm{sym}}$ and $L$ implies that one can extract
$K_{\mathrm{sym}}$ from $L$.

As mentioned above, an one-parameter parametrization for the
symmetry energy is sometimes used for simplicity, i.e.,
\begin{equation}
E_{\text{\textrm{sym}}}(\rho )=E_{\text{\textrm{sym}}}({\rho _{0}})\left(
\frac{{\rho }}{{\rho _{0}}}\right) ^{\gamma }  \label{EsymSimp1}
\end{equation}%
or%
\begin{eqnarray}
E_{\text{\textrm{sym}}}(\rho ) &=&E_{\text{\textrm{sym}}}^{\rm kin}({\rho _{0}}%
)\left( \frac{{\rho }}{{\rho _{0}}}\right) ^{2/3}  \notag \\
&&+\left[ E_{\text{\textrm{sym}}}({\rho _{0}})
-E_{\text{\textrm{sym}}}^{\rm kin}(%
{\rho _{0}})\right] \left( \frac{{\rho }}{{\rho _{0}}}\right) ^{\gamma }.
\label{EsymSimp2}
\end{eqnarray}%
For the parametrization of Eq. (\ref{EsymSimp1}), the $L$ and $K_{\mathrm{sym%
}}$ can be obtained as%
\begin{eqnarray}
L &=&3\gamma E_{\text{\textrm{sym}}}({\rho _{0}}) \\
K_{\text{\textrm{sym}}} &=&9\gamma (\gamma
-1)E_{\text{\textrm{sym}}}({\rho _{0}},)  \notag
\end{eqnarray}%
and furthermore we have%
\begin{equation}
K_{\text{\textrm{sym}}}=\frac{L^{2}}{E_{\text{\textrm{sym}}}({\rho
_{0}})}-3L, \notag
\end{equation}%
which indicates that the $K_{\text{\textrm{sym}}}$ parameter is
quadratically correlated with $L$. The result obtained by varying
the parameter $\gamma$ is shown in Fig. \ref{KsymL} by the dotted
line. Similarly, for the parametrization of Eq. (\ref{EsymSimp2}),
the $L$ and
$K_{\mathrm{sym}}$ can be expressed as%
\begin{eqnarray}
L &=&2E_{\text{\textrm{sym}}}^{\rm kin}({\rho _{0}})+3\gamma \left[ E_{\text{%
\textrm{sym}}}({\rho _{0}})-E_{\text{\textrm{sym}}}^{kin}({\rho _{0}})\right]
\\
K_{\text{\textrm{sym}}} &=&9\gamma (\gamma -1)\left[ E_{\text{\textrm{sym}}}(%
{\rho _{0}})-E_{\text{\textrm{sym}}}^{\rm kin}({\rho _{0}})\right]  \notag \\
&&-2E_{\text{\textrm{sym}}}^{\rm kin}({\rho _{0}}),
\end{eqnarray}%
and the $K_{\text{\textrm{sym}}}$ parameter is again quadratically
correlated with $L$, and this is illustrated by the dash-dot-dotted
line in Fig. \ref{KsymL} obtained by varying the value of $\gamma$.

It is very interesting to see from Fig. \ref{KsymL} that for larger
$L$ values ($L\geq 45$ MeV which is consistent with the constraint
from heavy-ion collision data shown later), all above symmetry
energy functionals from different models and parameterizations give
consistent predictions for the\ $K_{\mathrm{sym}}$-$L$ correlation.
This nice feature implies that using these different models and
parameterizations for the symmetry energy will not influence
significantly the determination of the $K_{\mathrm{sat,2}}$
parameter. On the other hand, the\ $K_{\mathrm{sym}}$-$L$
correlation from the two one-parameter parameterizations on the
symmetry energies in Eqs. (\ref{EsymSimp1}) and (\ref{EsymSimp2})
deviate significantly from the MDI, MSL and SHF predictions for
small $L$ values. Actually, the two forms of one-parameter
parametrization for the symmetry energy in Eqs. (\ref{EsymSimp1})
and (\ref{EsymSimp2}) may be too simple to describe a softer
symmetry energy. As shown in Ref. \cite{Che05a}, Eq.
(\ref{EsymSimp1}) cannot describe correctly the density dependence
of the symmetry energy from the MDI interaction with $x=1$ (the
Gogny interaction). On the other hand, as shown in Fig.
\ref{EsymRho}, the MSL model can give a nice description of the
density dependence of the symmetry energy from the very soft to the
very stiff. Although there is no direct empirical information on the
$K_{\mathrm{sym}}$ parameter and some uncertainty on the
$K_{\mathrm{sym}}$-$L$ correlation still exist, we assume here that
the correlation between $K_{\mathrm{sym}}$ and $L$ from the MSL
model with $\gamma _{\mathrm{sym}}=4/3 $ and $5/3$ is valid and then
use the experimental constraint on $L$ to extract the value of
$K_{\mathrm{sym}}$.

\subsubsection{Phenomenological MSL model constraint on the $K_{\mathrm{sat,2%
}}$ parameter}

\begin{figure}[tbh]
\includegraphics[scale=1.2]{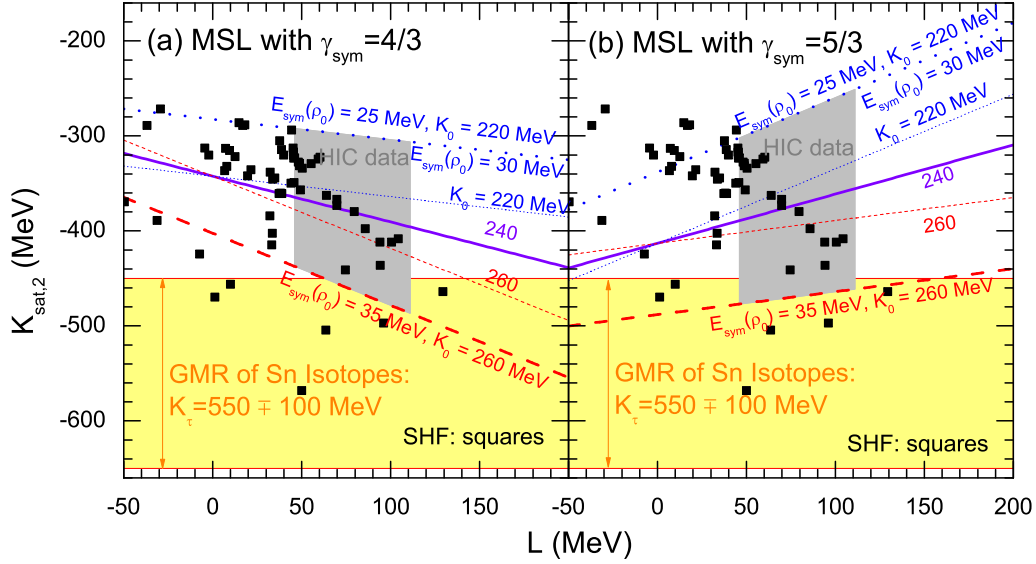}
\caption{{\protect\small (Color online) }$K_{\mathrm{sat,2}}${\protect\small %
\ as a function of }$L${\protect\small \ from the MSL model with }$\protect%
\gamma _{\mathrm{sym}}=4/3${\protect\small \ (a) and }$5/3${\protect\small \
(b) and }$m_{s,0}^{\ast }=0.8m${\protect\small \ and }$m_{v,0}^{\ast }=0.7m$%
{\protect\small \ for different values of }$K_{0}${\protect\small \ and }$E_{%
\text{\textrm{sym}}}(\protect\rho _{0})${\protect\small . The shaded region
indicates constraints within the MSL model with }$K_{0}=240\pm 20\ $%
{\protect\small MeV, }$E_{\text{\textrm{sym}}}(\protect\rho _{0})=30\pm 5$%
{\protect\small \ MeV, and }$46${\protect\small \ MeV }$\leq L\leq 111$%
{\protect\small \ MeV limited by the heavy-ion collision data. The
results from the SHF approach with 63 Skyrme interactions are also
included for
comparison. In addition, the constraint of }$K_{\protect\tau }=-550\pm 100$%
{\protect\small \ MeV obtained in Ref. \protect\cite{LiT07,Gar07} from
measurements of the isotopic dependence of the GMR in even-A Sn isotopes is
also indicated.}}
\label{Ksat2L}
\end{figure}
From above analyses, we can now extract information on the
$K_{\mathrm{sat,2}}$ parameter from the experimental constraints on
the $K_{0}$ parameter and the $L$ parameter within the
phenomenological MSL model. As pointed out previously, the value of
$K_{0}$ has been relatively well determined to be $240\pm 20$ MeV
from the nuclear GMR \cite{You99,Shl06,LiT07,Gar07,Col09}. The slope
parameter $L$ has been found to correlate linearly with the
neutron-skin thickness of heavy nuclei and thus can in principle be
determined from measured thickness of the neutron skin of such
nuclei \cite{Bro00,Hor01a,Typ01,Fur02,Kar02,Die03,Che05b,Ste05b}.
Unfortunately, because of the large uncertainties in the
experimental measurements, this has not yet been possible so far.
The proposed experiment of parity-violating electron scattering from
$^{208}$Pb, i.e., Parity Radius Experiment (PREx) at the Jefferson
Laboratory is expected to give an independent and accurate
measurement of its neutron skin thickness (within $0.05$ fm)
\cite{Hor01,Mic05} and thus to impose a stringent constraint on the
slope parameter $L$ in future. On the other hand, as mentioned
previously, heavy-ion collisions, especially those induced by
neutron-rich nuclei, provide a unique tool to explore the density
dependence of the symmetry energy and thus the $L$ parameter.
Actually, as discussed previously, the $L$ parameter has been
already limited significantly by heavy-ion collisions data.

From Eqs. (\ref{Ksat2}) and (\ref{KsymLMID}), we have in the MSL
model
\begin{eqnarray}
K_{\mathrm{sat,2}} &=&K_{\mathrm{sym}}-6L-\frac{J_{0}}{K_{0}}L  \notag \\
&=&-\left(\frac{J_{0}}{K_{0}}+6-3\gamma _{\mathrm{sym}}\right)L  \notag \\
&&+(3\gamma _{\mathrm{sym}}-2)E_{\text{\textrm{sym}}}^{\rm kin}({\rho _{0}}%
)+2D(5-3\gamma _{\mathrm{sym}})  \notag \\
&&-9\gamma _{\mathrm{sym}}E_{\text{\textrm{sym}}}({\rho _{0}}).
\label{Ksat2MID}
\end{eqnarray}%
Results on $K_{\mathrm{sat,2}}$ as a function of $L$ are shown in
Fig. \ref{Ksat2L} for $\gamma _{\mathrm{sym}}=4/3$ (panel (a)) and
$5/3$ (panel (b)) and with $K_{0}=220$, $240$, and $260$ MeV. In Eq.
(\ref{Ksat2MID}), the value of $J_{0}$ is obtained from Eq.
(\ref{J0}) using the value of $K_{0}$. For other quantities in the
MSL model, the default values $E_{\text{\textrm{sym}}}^{\rm
kin}({\rho _{0}})=12.3$ MeV, $E_{\text{\textrm{sym}}}({\rho
_{0}})=30$ MeV, and $D=-3.51$ MeV from $m_{s,0}^{\ast }=0.8m$ and
$m_{v,0}^{\ast }=0.7m$ have been used in the calculations. Since the
value of $2D(5-3\gamma _{\mathrm{sym}})$ is only about $7$ MeV for
$\gamma _{\mathrm{sym}}=4/3$ and is zero for
$\gamma_{\mathrm{sym}}=5/3$, as mentioned before, the contribution
of the $D$ parameter to $K_{\mathrm{sat,2}}$ is thus small. Also,
the correlation between $K_{\mathrm{sym}}$ and $L$ depends on
$E_{\text{\textrm{sym}}}({\rho _{0}})$ and $D$ as shown in Eq.
(\ref{KsymLMID}). To take into consideration of the uncertainty in
the value of $E_{\text{\textrm{sym}}}({\rho _{0}})$, we also include
in Fig. \ref{Ksat2L} the results with $K_{0}=220$ MeV and
$E_{\text{\textrm{sym}}}({\rho _{0}})=25 $ MeV as well as
$K_{0}=260$ MeV and $E_{\text{\textrm{sym}}}({\rho _{0}})=35 $ MeV,
which represent, respectively, the upper and lower bounds for a
fixed value of $L$. The shaded region in Fig. \ref{Ksat2L} further
shows the constrained $L$ values from heavy-ion collision data,
namely, $46$ MeV $\leq L\leq 111$ MeV. The lower limit of $L=46$ MeV
is obtained from the lower bound in the ImQMD analyses of the
isospin diffusion data and the double neutron/proton ratio
\cite{Tsa09} while the upper limit of $L=111$ MeV corresponds to the
upper bound of $L$ from the IBUU04 transport model analysis of the
isospin diffusion data \cite{Tsa04,Che05a,LiBA05c,Che05b}. The
constraint $46$ MeV $\leq L\leq 111$ MeV is consistent with the
analyses of the pygmy dipole resonances \cite{Kli07}, the giant
dipole resonance (GDR) of $^{208}$Pb analyzed with Skyrme
interactions \cite{Tri08}, the very precise Thomas-Fermi model fit
to the binding energies of $1654$ nuclei~\cite{Mye96}, and the
recent neutron-skin analysis \cite{Cen09}. These empirically
extracted values of $L$ represent the best and most stringent
phenomenological constraints available so far on the nuclear
symmetry energy at sub-saturation densities.

It is seen from Fig. \ref{Ksat2L} that the $K_{\mathrm{sat,2}}$
parameter decreases with increasing $L$ for $\gamma
_{\mathrm{sym}}=4/3$ while the opposite behavior is observed for
$\gamma _{\mathrm{sym}}=5/3$. This feature can be easily understood
from Eq. (\ref{Ksat2MID}). For $\gamma _{\mathrm{sym}}=4/3$, Eq.
(\ref{Ksat2MID}) is reduced to
\begin{equation}
K_{\mathrm{sat,2}}=-(\frac{J_{0}}{K_{0}}+2)L-12E_{\text{\textrm{sym}}}({\rho
_{0}})+17.6\text{ (MeV)},  \label{Ksat2L1}
\end{equation}%
while for $\gamma _{\mathrm{sym}}=5/3$, it is reduced to
\begin{equation}
K_{\mathrm{sat,2}}=-(\frac{J_{0}}{K_{0}}+1)L-15E_{\text{\textrm{sym}}}({\rho
_{0}})+36.9\text{ (MeV).}  \label{Ksat2L2}
\end{equation}%
For $K_{0}=240\pm 20$ MeV, Fig. \ref{J0K0abcSHFMDI} (or Eq.
(\ref{J0})) shows that the value of $J_{0}/K_{0}$ varies from about
$-1.9$ to $-1.2$ when $m_{s,0}^{\ast }$ changes from $m$ to $0.7m$.
Eq. (\ref{Ksat2L1}) (Eq. (\ref{Ksat2L2})) thus shows that
$K_{\mathrm{sat,2}} $ decreases (increases) with increasing $L$ for
$\gamma _{\mathrm{sym}}=4/3 $ ($5/3$).

An interesting feature observed from Fig. \ref{Ksat2L} is that the
$K_{\mathrm{sat,2}}$ parameter depends significantly on the symmetry
energy at the normal nuclear density $E_{\text{\textrm{sym}}}({\rho
_{0}})$. This can be seen more clearly from Eqs. (\ref{Ksat2L1}) and
(\ref{Ksat2L2}) which show that changing
$E_{\text{\textrm{sym}}}({\rho _{0}})$ by $5$ MeV leads to a
variation of $60-75$ MeV for $K_{\mathrm{sat,2}}$. This feature
indicates that an accurate determination of
$E_{\text{\textrm{sym}}}({\rho _{0}})$ is important for determining
the value of $K_{\mathrm{sat,2}}$. In addition, we have used
$E_{\text{\textrm{sym}}}({\rho _{0}})=30$ MeV in Fig. \ref{KsymL}
and one can easily see that a variation of $5$ MeV for
$E_{\text{\textrm{sym}}}({\rho _{0}})$ would lead to a shift of
$60-75$ MeV for the $K_{\text{\textrm{sym}}}$ parameter for a fixed
$L$. Therefore, the MSL model with $\gamma _{\mathrm{sym}}=4/3$ and
$5/3$ together with $E_{\text{\textrm{sym}}}({\rho _{0}})=30\pm 5$
MeV may provide a nice estimate for both the $K_{\mathrm{sym}}$-$L$
correlation and its uncertainty.

From the shaded region indicated in Fig. \ref{Ksat2L}, we find that
for $\gamma _{\mathrm{sym}}=4/3$, we have $-437$ MeV $\leq
K_{\mathrm{sat,2}}\leq -292$ MeV for $L=46$ MeV while $-487$ MeV
$\leq K_{\mathrm{sat,2}}\leq -306$ MeV for $L=111$ MeV. For $\gamma
_{\mathrm{sym}}=5/3$, we have $-477$ MeV $\leq
K_{\mathrm{sat,2}}\leq -302$ MeV for $L=46$ MeV while $-461$ MeV
$\leq K_{\mathrm{sat,2}}\leq -251$ MeV for $L=111$ MeV. These
results indicate that based on the MSL model with $4/3\leq \gamma
_{\mathrm{sym}}\leq 5/3$, $K_{0}=240\pm 20$ MeV, $25$ MeV $\leq
E_{\text{\textrm{sym}}}(\rho _{0})\leq 35$ MeV,\ and $46$ MeV $\leq
L\leq 111$ MeV, the $K_{\mathrm{sat,2}}$\ parameter can vary from
$-251$ MeV to $-487$ MeV. The results shown in Fig. \ref{Ksat2L} is
obtained from a $J_{0}/K_{0}$ value that is evaluated with the
default value $m_{s,0}^{\ast }=0.8m$. Similar analyses indicate that
the $K_{\mathrm{sat,2}}$\ parameter varies from $-261$ MeV to $-489$
MeV if we use $m_{s,0}^{\ast }=0.7m$ while it varies from $-245$ MeV
to $-485$ MeV if $m_{s,0}^{\ast }=0.9m $ is used. These results
indicate that the extracted value for $K_{\mathrm{sat,2}}$ is not
sensitive to the variation of the nucleon effective mass. The MSL
model analyses with $4/3\leq \gamma _{\mathrm{sym}}\leq 5/3$,
$K_{0}=240\pm 20$ MeV, $E_{\text{\textrm{sym}}}({\rho _{0}})=30\pm
5$ MeV, $46$ MeV $\leq L\leq 111$ MeV, and $m_{s,0}^{\ast }=0.8\pm
0.1m$ thus lead to an estimate of $K_{\mathrm{sat,2}}=-370\pm 120$
MeV.

Also included in Fig. \ref{Ksat2L} are the results from the SHF
approach with the $63$ Skyrme interactions. It is seen that among
the $63$ Skyrme interactions shown in Tables \ref{E0tab1},
\ref{E0tab2}, and \ref{E0tab3}, there are only $19$ Skyrme
interactions with predicted $K_{\mathrm{sat,2}}$ consistent with the
constraints obtained here (the shaded regions in Fig. \ref{Ksat2L}
(a) or (b)). The $19$ Skyrme interactions are SKa, SKT4, G$_{\sigma
}$, SkI3, SkI2, R$_{\sigma }$, SKO, SKO$^{\ast }$, SkMP, SGI, SKM,
SLy5, SLy1, SLy9, SLy2, SLy7, SkI6, SLy230b, and SkI4. From Tables
\ref{E0tab1} - \ref{Esymtab3}, one can see that all these $19$
Skyrme interactions predict $0.151$ fm$^{-3}$ $\leq \rho _{0}\leq $
$0.161$ fm$^{-3}$, $-16.0$ MeV $\leq E_{0}({\rho _{0}})\leq -15.6$
MeV, $217$ MeV $\leq K_{0}\leq 263$ MeV, $28$ MeV $\leq
E_{\text{\textrm{sym}}}({\rho _{0}})\leq 35$ MeV, and $46$ MeV $\leq
L\leq 104$ MeV, which are consistent with the empirical information.
However, the value of the nucleon effective mass predicted by the
$19$ Skyrme interactions varies widely, i.e., from $0.58m$ to $m$
for $m_{s,0}^{\ast }$ while from $0.51m$ to $m$ for $m_{v,0}^{\ast
}$. If we further impose the constraints of $m_{s,0}^{\ast }=0.8\pm
0.1m$ and $m_{s,0}^{\ast }>m_{v,0}^{\ast }$, then only five Skyrme
interactions, namely, G$_{\sigma }$, R$_{\sigma }$, SKM, SKO and
SKO$^{\ast }$ remain to be consistent with all known empirical
constraints except that the SKM interaction gives a little smaller
incompressibility $K_{0} = 216.6$ MeV than the empirical constrained
values of $K_{0}=240\pm 20$ MeV. These features imply that the
constraint on the $K_{\mathrm{sat,2}}$ parameter does not
significantly limit the nucleon effective mass as the former has
previously been shown to be insensitive to the latter.

\subsubsection{Discussions on
parameters $K_{\mathrm{sat,2}}$, $K_{\mathrm{asy}}$, and
$K_{\protect\tau }$}

As shown in Eq. (\ref{Ksat2Kasy}), the $K_{\mathrm{asy}}$ parameter
corresponds to the $K_{\mathrm{sat,2}}$ parameter when $J_{0}$ is
zero, i.e., the parabolic approximation to the EOS of symmetric
nuclear matter (Eq. (\ref{E0para})) is valid. In the MSL model, a
vanishing $J_{0}$ requires a $K_{0}$ value of about $340$ MeV, which
is significantly larger than the empirical value of $240\pm 20$ MeV.
Since $J_{0}/K_{0}\approx -1.5$ for $K_{0}=240$ MeV in the MSL
model, we have $K_{\mathrm{sat,2}}\approx K_{\mathrm{asy}}+1.5L$ or
$K_{\mathrm{asy}}\approx K_{\mathrm{sat,2}}-1.5L$. Therefore, the
difference between the $K_{\mathrm{asy}}$ parameter and the
$K_{\mathrm{sat,2}}$ parameter depends on the slope parameter $L$ of
the symmetry energy and a larger\ $L$ value (stiffer symmetry
energy) would lead to a larger difference. It should be stressed,
however, that the $K_{\mathrm{asy}}$ parameter is completely
determined by the density dependence of symmetry energy regardless
of the EOS\ of symmetric nuclear matter.

Based on the IBUU04 transport model analysis of the isospin
diffusion data, a value of $K_{\mathrm{asy}}=-500\pm 50$ MeV has
been extracted from the symmetry energy obtained by the MDI
interaction with the $x$ parameter between $0$ and $-1$. The
extracted $K_{\mathrm{asy}}=-500\pm 50$ MeV is essentially
consistent with the symmetry energy extracted from the ImQMD model
analyses of both the isospin diffusion data and the double
neutron/proton ratio \cite{Tsa09}, which predict a
$K_{\mathrm{asy}}$ value from about $-500 $ MeV to $-340$ MeV, and
the symmetry energy $E_{\mathrm{sym}}(\rho )=31.6(\rho /\rho
_{0})^{0.69}$ obtained from the isoscaling analyses \cite{She07},
which gives $K_{\mathrm{asy}}\approx -453$ MeV. Furthermore, the
constraint $K_{\mathrm{asy}}=-500\pm 50$ MeV is consistent with the
very recent constraint of $K_{\mathrm{asy}}\approx
-500_{-100}^{+125}$ MeV obtained from the study of neutron skin of
finite nuclei \cite{Cen09}. In the MDI interaction, one finds
$J_{0}/K_{0}=-2.1$ from Table \ref{E0tab1} and we thus have $-311$
MeV $\leq K_{\mathrm{sat,2}}\leq -316$ MeV, corresponding to the
prediction from the MDI interaction with the $x$ parameter between
$0$ and $-1$. Therefore, the $K_{\mathrm{sat,2}}$ value for the MDI
interaction is significantly larger than the $K_{\mathrm{asy}}$
value and is insensitive to the density dependence of symmetry
energy. On the other hand, the $J_{0}$ parameter in the MDI
interaction is important for the determination of the
$K_{\mathrm{sat,2}}$ parameter. The importance of the higher-order
$J_{0}$ parameter to $K_{\mathrm{sat,2}}$ can be further seen from
Fig. \ref{Ksat24KasyL}, which shows that for the $63$ Skyrme
interactions considered here, the magnitude of the
$K_{\mathrm{asy}}$ parameter is generally larger than that of the
$K_{\mathrm{sat,2}}$ parameter, especially, for the stiffer symmetry
energies (larger $L$ values). These results indicate that the
higher-order $J_{0}$ contribution to $K_{\mathrm{sat,2}}$ generally
cannot be neglected.

In Fig. \ref{Ksat2L}, the constraint $K_{\tau }=-550\pm 100$\ MeV
obtained in Ref. \cite{LiT07,Gar07} from recent measurements of the
isotopic dependence of the GMR in even-A Sn isotopes is also shown.
While the estimate of $K_{\mathrm{sat,2}}=-370\pm 120$ MeV obtained
in the present work has small overlap with the constraint of
$K_{\tau }=-550\pm 100$\ MeV, the latter still has significantly
larger magnitude than the former. According to Fig. \ref{Ksat2L},
there are only $6$ Skyrme interactions, namely, SII, SIV, SV, SI,
SkI5, SIII that predict $K_{\mathrm{sat,2}}=-550\pm 100$\ MeV. As
shown in Tables \ref{E0tab1}, \ref{E0tab2}, and \ref{E0tab3}, the
values for $K_{0}$ from SI, SII, SIII SIV, and SV are all larger
than $305$ MeV which are obviously inconsistent with the empirical
constraint of $K_{0}=240\pm 20$ MeV. For SkI5, we have $L=129.3$ MeV
which is significantly larger than the empirical constraint of $46$
MeV $\leq L\leq 111$ MeV. Therefore, none of the $63 $ Skyrme
interactions considered in the present work is consistent
simultaneously with $K_{\mathrm{sat,2}}=-550\pm 100$\ MeV and the
empirical constraints of $K_{0}=240\pm 20$ MeV and $46$ MeV $\leq
L\leq 111$ MeV. These features imply that the $K_{\tau }$ parameter
extracted from Eq. (\ref{KA2}) based on the GMR may not fully
reflect the $K_{\mathrm{sat,2}}$ parameter. As mentioned before, the
$K_{\tau }$ parameter may depend on the detailed truncation scheme
in Eq. (\ref{KA1}). The constraint $K_{\tau }=-550\pm 100$\ MeV
obtained in Ref. \cite{LiT07,Gar07} is based on Eq. (\ref{KA2}) and
thus neglects contributions from the $K_{\mathrm{curv}}$ term,
$K_{\mathrm{ss}}$  and other higher-order terms in Eq. (\ref{KA1}).
It is expected that including contributions from higher-order terms
in Eq. (\ref{KA2}) may change the extracted value for the $K_{\tau
}$ parameter as found in Ref. \cite{Shl93}.

\section{Summary and conclusions}

\label{summary}

We have studied in the present paper higher-order effects on the
properties of isospin asymmetric nuclear matter when the EOS and
saturation properties of asymmetric nuclear matter are expanded in
powers of the isospin asymmetry parameter $\delta =(\rho _{n}-\rho
_{p})/\rho $ and the dimensionless variable $\chi =(\rho -\rho
_{0})/3\rho _{0}$ that characterizes the deviations of the density
from the normal nuclear density $\rho _{0}$. \emph{Analytical
expressions for the saturation density of asymmetric nuclear matter
as well as  its binding energy and incompressibility at saturation
density have been derived and given exactly up to $4$th-order in the
isospin asymmetry $\delta =(\rho _{n}-\rho _{p})/\rho $} using $11$
characteristic parameters defined at the normal nuclear density
$\rho _{0}$ by the density derivatives of the binding energy per
nucleon of symmetric nuclear matter, the symmetry energy
$E_{\text{\textrm{sym}}}(\rho ) $ and the $4$th-order symmetry
energy $E_{\text{\textrm{sym,4}}}(\rho )$, namely, $E_{0}(\rho
_{0})$, $K_{0}$, $J_{0}$, $I_{0}$, $E_{\mathrm{sym}}(\rho _{0})$,
$L$, $K_{\mathrm{sym}}$, $J_{\mathrm{sym}}$,
$E_{\mathrm{sym,4}}(\rho _{0})$, $L_{\mathrm{sym,4}}$, and
$K_{\mathrm{sym,4}}$. Our method/recipe to derive the analytical
expressions for the saturation properties of asymmetry nuclear
matter is quite general, and in principle, the higher-order
coefficients (higher than $\delta ^{4}$) in the isospin asymmetry
$\delta $ can be easily obtained. Using the isospin- and
momentum-dependent MDI interaction, the SHF approach with $63$
popular Skyrme interactions, and the phenomenological MSL model, we
have systematically studied the higher-order effects on the
properties of asymmetric nuclear matter.

Firstly, our results indicate that including terms up to $\chi ^{4}$
can give a good description of the EOS of symmetric nuclear matter
$E_{0}(\rho )$, the density-dependent symmetry energy
$E_{\text{\textrm{sym}}}(\rho )$, and the density-dependent
$4$th-order symmetry energy $E_{\text{\textrm{sym,4}}}(\rho )$ for
densities less than about $2\rho _{0}$ while higher-order terms in
$\chi $ (higher than $4$th-order and thus more characteristic
parameters defined at $\rho _{0}$ are necessary) are generally
needed to describe reasonably the $E_{0}(\rho )$,
$E_{\text{\textrm{sym}}}(\rho )$, and
$E_{\text{\textrm{sym,4}}}(\rho )$ at higher density region (above
$2\rho _{0}$). \emph{These features imply that it is very difficult
to predict the high density behaviors of the EOS of asymmetric
nuclear matter based on the characteristic parameters obtained at
the normal nuclear density $\rho _{0}$}. Therefore, it is of great
interest to use the transport model analysis of heavy-ion collisions
at intermediate and high energies as well as the astrophysical
observations of compact stars as tools to extract information on the
EOS of asymmetric nuclear matter at high densities.

Secondly, we have systematically studied the isospin dependence of
the saturation properties of asymmetric nuclear matter, i.e., the
saturation density $\rho _{\mathrm{sat}}(\delta )$ as well as the
binding energy $E_{\mathrm{sat}}(\delta )$ and incompressibility
$K_{\mathrm{sat}}(\delta )$ at saturation density. In particular, we
have compared their exact results with the analytical expressions
expanded up to the $\delta ^{2}$ term and the $\delta ^{4}$ term,
respectively. Our results indicate that the saturation properties of
asymmetric nuclear matter, i.e., $\rho _{\mathrm{sat}}(\delta ) $,
$E_{\mathrm{sat}}(\delta )$ and $K_{\mathrm{sat}}(\delta )$, exhibit
a good linear dependence on $\delta ^{2}$ at smaller isospin
asymmetries with $\delta ^{2}\leq 0.1$ which is relevant to the
structure of finite heavy nuclei. \emph{This feature implies that
the higher-order terms (the $\delta ^{4}$ term and higher-order
terms in $\delta $) are not important for the saturation properties
of asymmetric nuclear matter, at least for an asymmetric nuclear
matter that is not extremely neutron-rich (proton-rich).} On the
other hand, for asymmetric nuclear matter with extremely large
isospin asymmetries with $\delta ^{2}\geq 0.3$ and depending on the
stiffness of the nuclear symmetry energy, the higher-order terms in
$\delta $ ($\delta^{4}$ and higher-order terms) may become important
for describing reasonably the saturation properties. It is further
found that the saturation density and the incompressibility at
saturation density generally decrease with the magnitude of the
isospin asymmetry while the binding energy at saturation density
shows the opposite behavior.

Finally, we have studied in detail the second-order isospin
coefficient $K_{\mathrm{sat,2}}$ of the incompressibility of an
asymmetric nuclear matter at its saturation density. It is found
that the magnitude of the higher-order $K_{\mathrm{sat,4}}$
parameter is generally small compared to that of the
$K_{\mathrm{sat,2}}$ parameter, so the latter essentially
characterizes the isospin dependence of the incompressibility at
saturation density. Furthermore, we have found that the
$K_{\mathrm{sat,2}}$ parameter is uniquely determined by $L$,
$K_{\mathrm{sym}}$ and $J_{0}/K_{0}$. Since there is no experimental
information on the $J_{0}$ parameter and the $K_{\mathrm{sym}}$
parameter, we have thus used the MSL model, which can reasonably
describe the general properties of symmetric nuclear matter and the
symmetry energy predicted by both the MDI model and the SHF
approach, to estimate the value of $K_{\mathrm{sat,2}}$. \emph{Our
results indicate that generally the higher-order $J_{0}$
contribution to $K_{\mathrm{sat,2}}$ cannot be neglected, especially
for larger $L$ values. Interestingly, it is found that there exists
a nicely linear correlation between $K_{\mathrm{sym}}$ and $L$ as
well as between $J_{0}/K_{0}$ and $K_{0}$} for the three different
models used here, i.e., the MDI interaction, the MSL interaction,
and the SHF approach with $63$ Skyrme interactions. For the MSL
model, the correlation between $K_{\mathrm{sym}}$ and $L$ is further
found to depend significantly on the value of
$E_{\text{\textrm{sym}}}(\rho _{0})$ but not on variations of the
nucleon effective mass. These correlations and features have allowed
us to extract the values of the $J_{0}$ parameter and the
$K_{\mathrm{sym}}$ parameter from the empirical information on
$K_{0}$, $L$ and $E_{\text{\textrm{sym}}}(\rho _{0})$. \emph{In
particular, using the empirical constraints of $K_{0}=240\pm 20$
MeV, $E_{\text{\textrm{sym}}}({\rho _{0}})=30\pm 5$ MeV, $46 $ MeV
$\leq L\leq 111$ MeV and $m_{s,0}^{\ast }=0.8\pm 0.1m$ in the MSL
model leads to an estimate of $K_{\mathrm{sat,2}}=-370\pm 120$ MeV.}

While the estimated value of $K_{\mathrm{sat,2}}=-370\pm 120$ MeV in
the present work has small overlap with the constraint of $K_{\tau
}=-550\pm 100$ MeV obtained in Refs. \cite{LiT07,Gar07} from recent
measurements of the isotopic dependence of the GMR in even-A Sn
isotopes, its magnitude is significantly smaller than that of the
constrained $K_{\tau }$. Recently, there are several studies
\cite{Sag07,Pie07,Pie09} on extracting the value of the
$K_{\mathrm{sat,2}}$ parameter based on the idea initiated by
Blaizot and collaborators that the values of both $K_{0}$ and
$K_{\mathrm{sat,2}}$ should be extracted from the same consistent
theoretical model that successfully reproduces the experimental GMR
energies of a variety of nuclei. These studies show that no single
model (interaction) can simultaneously describe correctly the recent
measurements of the isotopic dependence of the GMR in even-A Sn
isotopes and the GMR data of $^{90}$Zr and $^{208}$Pb nuclei, and
this makes it difficult to accurately determine the value of
$K_{\mathrm{sat,2}}$ from these experimental data. Also, a very
recent study \cite{Kha09} indicates that the effect due to the
nuclear superfluidity may also affect the extraction of the
$K_{\mathrm{sat,2}}$ parameter from the nuclear GMR. As pointed out
in \cite{Pie09}, these features suggest that the $K_{\tau }=-550\pm
100$\ MeV obtained in Ref. \cite{LiT07,Gar07} may suffer from the
same ambiguities already encountered in earlier attempts to extract
the $K_{0}$ and $K_{\mathrm{sat,2}}$ of infinite nuclear matter from
finite-nuclei extrapolations. This problem remains an open
challenge, and both experimental and theoretical insights are needed
in future studies.

\begin{acknowledgments}
We thank A. Rios for useful communications. This work was supported
in part by the National Natural Science Foundation of China under
Grant Nos. 10575071 and 10675082, MOE of China under project
NCET-05-0392, Shanghai Rising-Star Program under Grant No.
06QA14024, the SRF for ROCS, SEM of China, the National Basic
Research Program of China (973 Program) under Contract No.
2007CB815004, the U.S. National Science Foundation under Grant No.
PHY-0652548, PHY-0757839 and PHY-0758115, the Welch Foundation under
Grant No. A-1358, the Research Corporation under Award No. 7123 and
the Texas Coordinating Board of Higher Education Award No.
003565-0004-2007.
\end{acknowledgments}

\appendix*

\section{Derivation of analytical expressions for the
saturation density and incompressibility to higher-order in isospin
asymmetry}

\label{AppendA}

For completeness, we present in this appendix a simple derivation of
the analytical expressions for the saturation density of asymmetric
nuclear matter and its incompressibility at saturation density up to
higher-order terms in the isospin asymmetry $\delta =(\rho _{n}-\rho
_{p})/\rho $. In particular, analytical expressions for the
saturation density and incompressibility at saturation density are
given exactly up to $4$th-order in $\delta $.

The binding energy per nucleon $e(\rho ,\delta )$ of asymmetric
nuclear matter can be expanded up to $2n$th-order in $\delta $ as
\begin{equation}
e(\rho ,\delta )=e_{0}(\rho )+e_{2}(\rho )\delta ^{2}+e_{4}(\rho
)\delta ^{4}+\cdots +e_{2n}(\rho )\delta ^{2n}. \label{EOSapp}
\end{equation}%
The first term $e_{0}(\rho )$ represents the EOS\ of symmetry
nuclear matter with $\delta =0$. We further expand each $e_{i}(\rho
)$ in Eq. (\ref{EOSapp}) up to $n$th-order in a dimensionless
variable $z=(\rho -\rho _{0})/\rho _{0}$, i.e.
\begin{equation}
\begin{array}{l}
e_{0}(\rho )=e_{0}(\rho _{0})+a_{01}z+a_{02}z^{2}+\cdots +a_{0n}z^{n} \\
e_{2}(\rho )=e_{2}(\rho _{0})+a_{11}z+a_{12}z^{2}+\cdots +a_{1n}z^{n} \\
e_{4}(\rho )=e_{4}(\rho _{0})+a_{21}z+a_{22}z^{2}+\cdots +a_{2n}z^{n} \\
\vdots \\
e_{2n}(\rho )=e_{2n}(\rho _{0})+a_{n1}z+a_{n2}z^{2}+\cdots
+a_{nn}z^{n}.%
\end{array}
\label{ChPAapp}
\end{equation}%
The coefficients $a_{ij}$ represent the characteristic parameters of
asymmetric nuclear matter defined at the normal nuclear density
$\rho _{0}$. The saturation density $\rho _{\mathrm{sat}}$ is then
determined by the following equation
\begin{equation}
\frac{\partial e}{\partial \rho }|_{\rho =\rho _{\mathrm{sat}}}=0
\label{rhoSATEqapp}
\end{equation}%
or equivalently%
\begin{equation}
\frac{\partial e}{\partial z}|_{z=z_{\mathrm{sat}}}=0
\label{zSATEqapp}
\end{equation}%
with $z_{\mathrm{sat}}=(\rho _{\mathrm{sat}}-\rho _{0})/\rho _{0}$

Substituting Eq. (\ref{EOSapp}) and Eq. (\ref{ChPAapp}) into Eq. (\ref%
{zSATEqapp}) leads to the following equation
\begin{equation}
\begin{array}{l}
a_{01}+2a_{02}z+3a_{03}z^{2}+\cdots +na_{0n}z^{n-1}+ \\
+(a_{11}+2a_{12}z+3a_{13}z^{2}+\cdots +na_{1n}z^{n-1})\delta ^{2}+ \\
+(a_{21}+2a_{22}z+3a_{23}z^{2}+\cdots +na_{2n}z^{n-1})\delta ^{4}+ \\
+\cdots + \\
+(a_{n1}+2a_{n2}z+3a_{n3}z^{2}+\cdots +na_{nn}z^{n-1})\delta
^{2n}=0.%
\end{array}
\label{xSATapp}
\end{equation}%
Obviously, we have $a_{01}=0$ as the saturation density of symmetry
nuclear matter is determined by $\frac{\partial e}{\partial \rho
}|_{\delta =0,\rho =\rho _{0}}=0$. In order to obtain the solution
to Eq. (\ref{xSATapp}), i.e.,\ $z_{\mathrm{sat}}$ with precision of
$O(\delta ^{2k+2})$, we assume
\begin{equation}
z_{\mathrm{sat}}=A_{2}\delta ^{2}+A_{4}\delta ^{4}+\cdots
+A_{2k}\delta ^{2k}, \label{zSATapp}
\end{equation}%
where $A_{2m}$ ($m=1,2,...,k)$ are coefficients to be determined
from comparing the coefficients of $\delta ^{2j}$ terms
($j=1,2,...,m)$ on the two sides of Eq. (\ref{xSATapp}) after
substituting Eq. (\ref{zSATapp}) into Eq. (\ref{xSATapp}). For
example, for $k=1$, i.e., $z_{\mathrm{sat}}=A_{2}\delta^{2}$, the
coefficient $A_{2}$ can be determined by the following equality
\begin{eqnarray}
&&2a_{02}A_{2}  \notag \\
&&+a_{11}=0.  \label{delta2cofapp}
\end{eqnarray}%
Similarly, for $k=2$, i.e., $z_{\mathrm{sat}}=A_{2}\delta ^{2}+A_{4}\delta
^{4}$, the coefficient $A_{4}$ can be determined by%
\begin{eqnarray}
&&2a_{02}A_{4}+3a_{03}A_{2}^{2}  \notag \\
&&+2a_{12}A_{2}  \notag \\
&&+a_{21}=0.  \label{delta4cofapp}
\end{eqnarray}%
Furthermore, for $k=3$ and $4$, the coefficients $A_{6}$ and $A_{8}$
can be determined, respectively, by
\begin{eqnarray}
&&2a_{02}A_{6}+6a_{03}A_{2}A_{4}+4a_{04}A_{2}^{3}  \notag \\
&&+2a_{12}A_{4}+3a_{13}A_{2}^{1}  \notag \\
&&+2a_{22}A_{2}  \notag \\
&&+a_{31}=0  \label{delta6cofapp}
\end{eqnarray}%
and%
\begin{eqnarray}
&&2a_{02}A_{8}+3a_{03}(2A_{2}A_{6}+A_{4}^{2})+12a_{04}A_{2}^{2}A_{4}+5a_{05}A_{2}^{4}
\notag \\
&&+2a_{12}A_{6}+6a_{13}A_{2}A_{4}+4a_{14}A_{2}^{3}  \notag \\
&&+2a_{22}A_{4}+3a_{23}A_{2}^{2}  \notag \\
&&+2a_{32}A_{2}  \notag \\
&&+a_{41}=0.  \label{delta8cofapp}
\end{eqnarray}%
From above analyses, the coefficient of the $\delta ^{2m}$ term,
$A_{2m}$, can be determined by all lower-order coefficients
$A_{2},A_{4},A_{2m-2}$ $(m=1,2,3,\ldots ,n)$. The $A_{2m}$ obtained
above are complete and precise up to the order of $\delta ^{2m}$,
and the higher-order characteristic parameters $a_{ij}$ do not
contribute to $A_{2m}$. In order to write down the general
expressions for $A_{2m}$, we define the following symbol
\begin{equation}
B_{n,m}=\sum {A_{2}^{i_{1}}A_{4}^{i_{2}}}\cdots A_{2n}^{i_{n}},
\label{eq9}
\end{equation}%
where we have $i_{1},i_{2},\ldots i_{n}\in \{0,N_{+}\}$ and they
satisfy following conditions
\begin{equation}
\sum\limits_{j=1}^{n}{ji_{j}=m,\qquad\sum\limits_{j=1}^{n}{i_{j}=n}}.
\label{eq10}
\end{equation}%
Then the coefficient of the $\delta ^{2m}$ term satisfies the
equation
\begin{equation}
\sum\limits_{j=0}^{m-1}{\sum\limits_{i=2}^{m-j+1}{ia_{ji}B_{{i-1},{m-j}}}}%
+a_{m1}=0  \label{eq11}
\end{equation}%
from which we can obtain the coefficient $A_{2m}$ as
\begin{eqnarray}
A_{2m} &=&-\frac{1}{a_{02}}\bigg(\sum\limits_{i=3}^{m+1}{ia_{0i}B_{{i-1},m}}
\notag \\
&&+\sum\limits_{j=1}^{m-1}{\sum\limits_{i=2}^{m-j+1}{ia_{ji}B_{{i-1},{m-j}}}%
+a_{m1}}\bigg).
\end{eqnarray}%
Therefore, one can obtain the saturation density of asymmetry
nuclear matter to any order of $\delta $.

Taking the precision to the order of $\delta ^{4}$, we then have
\begin{equation}
\begin{array}{l}
A_{2}=-\frac{a_{11}}{2a_{02}}, \\
A_{4}=\frac{a_{12}a_{11}}{2a_{02}^{2}}-\frac{3a_{03}a_{11}^{2}}{8a_{02}^{2}}-%
\frac{a_{21}}{2a_{02}}.
\end{array}
\label{eq13}
\end{equation}%
Converting the coefficients $a_{ij}$ into the conventional forms, i.e.,
\begin{equation}
\begin{array}{l}
a_{02}=\frac{K_{0}}{18},a_{03}=\frac{J_{0}}{162},a_{04}=\frac{I_{0}}{%
24\times 81}, \\
a_{11}=\frac{L}{3},a_{12}=\frac{K_{\mathrm{sym}}}{18},a_{13}=\frac{J_{%
\mathrm{sym}}}{162}, \\
a_{21}=\frac{L_{\mathrm{sym,4}}}{3},a_{22}=
\frac{K_{\mathrm{sym,4}}}{18}.%
\end{array}
\label{eq14}
\end{equation}%
we then obtain
\begin{equation}
z_{\mathrm{sat}}=-\frac{3L}{K_{0}}\delta ^{2}+\left( \frac{3K_{\mathrm{sym}}L%
}{K_{0}^{2}}-\frac{3L_{\mathrm{sym,4}}}{K_{0}}-\frac{3J_{0}L^{2}}{2K_{0}^{3}}%
\right) \delta ^{4}.  \label{eq15}
\end{equation}%
So the saturation density can be obtained as
\begin{eqnarray}
\rho _{\mathrm{sat}} &=&\rho _{0}\bigg[1-\frac{3L}{K_{0}}\delta ^{2}
\label{rhoSATapp} \\
&&+\left( \frac{3K_{\mathrm{sym}}L}{K_{0}^{2}}-\frac{3L_{\mathrm{sym,4}}}{%
K_{0}}-\frac{3J_{0}L^{2}}{2K_{0}^{3}}\right) \delta ^{4}+O(\delta
^{6})\bigg], \notag
\end{eqnarray}%
which is exactly Eq. (\ref{rhosat0}).

The incompressibility coefficient $K_{\mathrm{sat}}$ of asymmetry
nuclear matter at the saturation density is defined as
\begin{equation}
K_{\mathrm{sat}}=9\rho _{\mathrm{sat}}^{2}\frac{\partial ^{2}e}{\partial
\rho ^{2}}|_{\rho =\rho _{\mathrm{sat}}}.  \label{Kapp}
\end{equation}%
Substituting Eq. (\ref{EOSapp}) and Eq. (\ref{ChPAapp}) into Eq. (\ref{Kapp}%
) and using Eq. (\ref{rhoSATapp}), we can easily obtain
\begin{eqnarray}
K_{\mathrm{sat}} &=&9\rho _{0}^{2}\left[ 1-\frac{3L}{K_{0}}\delta
^{2}+\left( \frac{3K_{\mathrm{sym}}L}{K_{0}^{2}}-\frac{3L_{\mathrm{sym,4}}}{%
K_{0}}-\frac{3J_{0}L^{2}}{2K_{0}^{3}}\right) \delta ^{4}\right] ^{2}  \notag
\\
&&\times \bigg[\frac{K_{0}}{9\rho _{0}^{2}}+\frac{J_{0}}{27\rho _{0}^{2}}%
\frac{\rho -\rho _{0}}{\rho _{0}}+\frac{I_{0}}{162}\left( \frac{\rho -\rho
_{0}}{\rho _{0}}\right) ^{2}  \notag \\
&&+\left( \frac{K_{\mathrm{sym}}}{9\rho _{0}^{2}}+\frac{J_{\mathrm{sym}}}{%
27\rho _{0}^{2}}\frac{\rho -\rho _{0}}{\rho _{0}}+\frac{I_{\mathrm{sym}}}{162%
}\left( \frac{\rho -\rho _{0}}{\rho _{0}}\right) ^{2}\right) \delta ^{2}
\notag \\
&&+\left( \frac{K_{\mathrm{sym,4}}}{9\rho _{0}^{2}}+\frac{J_{\mathrm{sym,4}}%
}{27\rho _{0}^{2}}\frac{\rho -\rho _{0}}{\rho _{0}}\right) \delta
^{4}+O(\delta ^{6})\bigg]|_{\rho =\rho _{\mathrm{sat}}}  \notag \\
&=&K_{0}+\left( K_{\mathrm{sym}}-6L-\frac{J_{0}}{K_{0}}L\right) \delta ^{2}
\notag \\
&&+\bigg(K_{\mathrm{sym,4}}-6L_{\mathrm{sym,4}}-\frac{J_{0}L_{\mathrm{sym,4}}%
}{K_{0}}+\frac{9L^{2}}{K_{0}}-\frac{J_{\mathrm{sym}}L}{K_{0}}  \notag \\
&&+\frac{I_{0}L^{2}}{2K_{0}^{2}}+\frac{J_{0}K_{\mathrm{sym}}L}{K_{0}^{2}}+%
\frac{3J_{0}L^{2}}{K_{0}^{2}}-\frac{J_{0}^{2}L^{2}}{2K_{0}^{3}}\bigg)\delta
^{4}  \notag \\
&&+O(\delta ^{6}),
\end{eqnarray}%
which is exactly Eq. (\ref{Ksat0}). We note that the expression
$K_{\mathrm{sat}}=K_{0}+\left(
K_{\mathrm{sym}}-6L-\frac{J_{0}}{K_{0}}L\right) \delta ^{2}$ was
originally given in Ref. \cite{Bla80}. For terms higher than
$\delta^4$, they can be straightforwardly obtained following above
derivation but many more higher-order characteristic parameters
$a_{ij}$ would be needed.

\end{document}